\documentclass{aa}  
\usepackage{natbib,twoopt}
\usepackage{comment}
\usepackage[rightcaption]{sidecap}
\usepackage{hyperref} 
\hypersetup{
    colorlinks=true,
    linkcolor=red,
    filecolor=cyan,      
    urlcolor=magenta,
    pdftitle={Overleaf Example},
    pdfpagemode=FullScreen,
    citecolor=blue,
    breaklinks=true
    }
\bibpunct{(}{)}{;}{a}{}{,}             
\makeatletter
  \newcommandtwoopt{\citeads}[3][][]{\href{http://adsabs.harvard.edu/abs/#3}%
    {\def\hyper@linkstart##1##2{}%
     \let\hyper@linkend\@empty\citealp[#1][#2]{#3}}}
  \newcommandtwoopt{\citepads}[3][][]{\href{http://adsabs.harvard.edu/abs/#3}%
    {\def\hyper@linkstart##1##2{}%
     \let\hyper@linkend\@empty\citep[#1][#2]{#3}}}
  \newcommandtwoopt{\citetads}[3][][]{\href{http://adsabs.harvard.edu/abs/#3}%
    {\def\hyper@linkstart##1##2{}%
     \let\hyper@linkend\@empty\citet[#1][#2]{#3}}}
  \newcommandtwoopt{\citeyearads}[3][][]%
    {\href{http://adsabs.harvard.edu/abs/#3}
    {\def\hyper@linkstart##1##2{}%
     \let\hyper@linkend\@empty\citeyear[#1][#2]{#3}}}
\makeatother

\usepackage{adjustbox}
\usepackage{xcolor}
\usepackage{graphicx}
\usepackage{txfonts}
\usepackage{tabularx, blindtext}
\usepackage{rotating}
\usepackage{pdflscape}
\usepackage{inputenc}
\usepackage{caption}
\usepackage{longtable}
\usepackage{lscape}
\DeclareCaptionFormat{continued}{#1 (continued)#2#3\par}
\DeclareGraphicsExtensions{.jpg,.png, .pdf}
\usepackage[utf8]{inputenc}
\usepackage{xcolor}       
\usepackage{orcidlink}    

\definecolor{orcidlogocol}{HTML}{A6CE39}

\authorrunning{E. R. Garro et al.}
\titlerunning{The globular cluster Patchick~126}

\newcommand{\ESOChile}{ESO - European Southern Observatory, Alonso de Cordova 3107, Vitacura, Casilla 19001, Santiago, Chile}
\newcommand{\INAFBO}{INAF - Osservatorio di Astrofisica e Scienza dello Spazio di Bologna, Via Gobetti 93/3, 40129 Bologna, Italy}
\newcommand{\UCNa}{Universidad Cat\'olica del Norte, N\'ucleo UCN en Arqueolog\'ia Gal\'actica, Av. Angamos 0610, Antofagasta, Chile}
\newcommand{\UCNb}{Universidad Cat\'olica del Norte, Departamento de Ingenier\'ia de Sistemas y Computaci\'on, Av. Angamos 0610, Antofagasta, Chile}

\newcommand{\VATICAN}{Vatican Observatory, V-00120 Vatican City State, Italy}
\newcommand{\UNIBO}{Dipartimento di Fisica e Astronomia, Università degli Studi di Bologna, Via Gobetti 93/2, 40129 Bologna, Italy}
\newcommand{\McDonald}{Department of Astronomy and McDonald Observatory, The University of Texas, Austin, TX 78712, USA}
\newcommand{\EGEU}{Department of Astronomy and Space Sciences, Ege University, 35100 Bornova, {\. I}zmir, T{\" u}rkiye}
\newcommand{\atlanTTic}{atlanTTic, Universidade de Vigo, Escola de Enxeñaría de Telecomunicación, 36310, Vigo, Spain}
\newcommand{\uniLaLaguna}{Universidad de La Laguna, Avda. Astrofísico Fco. Sánchez, E-38205 La Laguna, Tenerife, Spain}

\begin{document} 
\title{HST+IGRINS synergy to characterise the newly discovered metal-rich bulge globular cluster Patchick 126}
   \author{
        Elisa R. Garro\inst{1}\orcidlink{0000-0002-4014-1591}
        \and
        Davide Massari\inst{2}\orcidlink{0000-0001-8892-4301}
        \and 
        Jos\'e G. Fern\'andez-Trincado\inst{3,4}\orcidlink{0000-0003-3526-5052}
        \and
        Edoardo Ceccarelli\inst{2,5}\orcidlink{0009-0007-3793-9766}
        \and
        Chris Sneden\inst{6}\orcidlink{0000-0002-3456-5929}
        \and
        Fernando Aguado-Agelet\inst{7,8}\orcidlink{0000-0003-2713-1943}
        \and
        Melike Af\c{s}ar\inst{9}\orcidlink{0000-0002-2516-1949}
        \and
        Michele Bellazzini\inst{2}\orcidlink{0000-0001-8200-810X}
        \and
        Rafael Guer\c{c}o\inst{3}
        \and
        Dante Minniti\inst{10,11}\orcidlink{0000-0002-7064-099X}
        \and 
        Mattia Libralato\inst{12}\orcidlink{0000-0001-9673-7397}
        \and
        Beatriz Barbuy\inst{13}\orcidlink{0000-0001-9264-4417}
        \and
        Bruno Dias\inst{10}
        }
   \institute{
          \ESOChile \\
         \email{elisaritagarro1@gmail.com} 
  \and
         \INAFBO
   \and
        \UCNa
   \and 
        \UCNb
   \and
        \UNIBO
 \and  
        \McDonald
 \and
     \atlanTTic
 \and
    \uniLaLaguna 
 \and
       \EGEU
  \and
 Instituto de Astrofísica, Depto. de Ciencias Físicas, Facultad de Ciencias Exactas, Universidad Andres Bello, Av. Fernandez Concha 700, Las Condes, Santiago, Chile
   \and
    \VATICAN
    \and
    INAF - Osservatorio Astronomico di Padova, Vicolo dell’Osservatorio 5, Padova I-35122, Italy
    \and
    Universidade de Sao Paulo, IAG, Rua do Matao 1226, Cidade Universitária, Sao Paulo 05508-900, Brazil
}
  \date{Received: 18 November 2025; Accepted: 23 December 2025}
 
\abstract
{ We present the first comprehensive spectroscopic and deep photometric study of the globular cluster candidate Patchick~126. 
The spectroscopic analysis is based on high-resolution near-infrared data obtained with the Immersion GRating INfrared Spectrometer (IGRINS) spectrograph, while the photometric analysis relies on Hubble Space Telescope (HST) observations from the \textit{Hubble Missing Globular Cluster Survey} (MGCS). 
We derived abundances for $\alpha$-(O, Mg, Si, Ca, Ti), light-(C, N), odd-Z (Na, Al), iron-peak (Fe, Co, Cr, Ni, Mn, V), and $s$-process elements (Ce) for four red giant stars observed in both the H and K bands. 
Our results yield a mean metallicity of $\langle\mathrm{[Fe/H]}\rangle = -0.30 \pm 0.03$, with no evidence of intrinsic variation, and an $\alpha$-enhancement of $\langle\mathrm{[\alpha/Fe]}\rangle = +0.19 \pm 0.02$, consistent with the trends of metal-rich Galactic globular clusters. 
We detect an intrinsic C--N anti-correlation, but no Na-O or Al-Mg anti-correlations, in agreement with expectations for low-mass, metal-rich clusters. 
From the HST photometry in the F606W and F814W bands, we constructed deep colour-magnitude diagrams extending $\sim 2$-$3$ magnitudes below the main-sequence turn-off. 
This depth allowed us to provide the first robust age estimate for the cluster. 
Applying the methods developed within the Cluster Ages to Reconstruct the Milky Way Assembly (CARMA) project, we derive an age of $11.9^{+0.3}_{-0.4}$~Gyr. 
Independently, we obtain a photometric metallicity of [Fe/H] $= -0.28$, in excellent agreement with the spectroscopic results. 
The colour excess we derived, E(B-V) = 1.08, confirms that Patchick~126 is a heavily reddened cluster, located at a heliocentric distance of $7.8$~kpc. 
Finally, from the orbital parameters, including energy, vertical angular momentum, circularity, and maximum vertical height, we find that Patchick~126 closely follows a disc-like orbit. Taken together, these results confirm that Patchick 126 is an in situ, low-mass globular cluster of the Milky Way, exhibiting properties that lie at the boundary between old-open and globular clusters.
}
   \maketitle

\section{Introduction}
\label{sec:Introduction}
Recently, numerous new globular cluster (GC) candidates have been identified in the obscured regions of the Milky Way (MW) bulge and disc, as reported in \cite{2024A&A...687A.214G} and \cite{2024A&A...687A.201B} compilations. These discoveries could increase the known GC population in the MW by up to $\sim20\%$. However, the vast majority of these objects remain classified as ‘candidates’, as their identification is based solely on photometric analyses combining near-infrared and optical datasets, such as the VISTA Variables in the Via Láctea Extended Survey (VVVX), \textit{Gaia} DR3, and the Two Micron All Sky Survey (2MASS), without spectroscopic confirmation. Only a few examples, such as VVV-CL001, Gran 1, 2, 3 (also known as Patchick 125), 4, 5, Garro 01, and LP 866, have been spectroscopically studied and reported in the recent literature (see e.g. \citealt{2022MNRAS.513.3993O, 2022A&A...657A..84F, 2023MNRAS.526.1075P, 2024A&A...683A.167G, 2024A&A...692A..14L}).\\

The available colour-magnitude diagrams (CMDs) for these star clusters, derived from optical and near-infrared data, primarily trace evolved evolutionary sequences, such as the red clump (RC), horizontal branch (HB), and red giant branch (RGB), since the main-sequence turn-off and lower main-sequence stars lie below the detection limits of these surveys \citep{2022A&A...662A..95G,2023A&A...669A.136G, 2024A&A...688L...3G}. As a result, although parameters such as reddening, distance, proper motions, and metallicity can be estimated, determining cluster age remains highly challenging. Furthermore, the final decontaminated CMDs reveal that many of these clusters are extremely faint ($M_{V}\approx 5$ mag, in the tail of the GC luminosity function, \citealt{2024A&A...687A.214G}) and sparse, hosting only a few stars with very low masses ($M\sim 10^{3} -10^{4}\ M_{\odot}$). These characteristics complicate their classification, leaving open whether they are truly GCs, low-mass open clusters, or intermediate-age objects -- whether formed in situ or ex situ, thus associated with past accretion events (see e.g. \citealt{2010MNRAS.404.1203F, 2019A&A...630L...4M, 2024A&A...691A.226C, 2024MNRAS.528.3198B, 2025arXiv250411687A}). For these reasons, our main long-term goal is to collect high-resolution spectroscopic data for most of these clusters in order to homogeneously characterise them. \\

From a spectroscopic point of view, GC stars have been known to exhibit inhomogeneities in their light-element abundances (C, N, O, Na, Mg, and/or Al), most notably as the well-established O--Na and Mg--Al anti-correlations \citep{2004ARA&A..42..385G,2010A&A...516A..55C, Bastian_2018}. For instance, the O--Na anti-correlation has been detected in many Galactic GCs \citep{2009A&A...505..139C} and is often considered a defining characteristic of these systems. Nevertheless, notable exceptions exist -- such as Ruprecht 106 \citep{Villanova_2013}, Palomar 12 \citep{Cohen_2004}, and Terzan 7 \citep{Sbordone_2005} -- where the trend is absent. In general, accreted clusters as well as low-mass and metal-rich systems tend to show little or no evidence of these anti-correlations \citep{Pancino_2017}. This behaviour is likely due to the self-enrichment that GCs experience during the early phases of their formation, which also enables the formation of multiple populations within the system \citep{2004ARA&A..42..385G}. Low- and intermediate-mass asymptotic giant branch (AGB) stars \citep{Ventura_2016,10.1093/mnras/stw387}, fast-rotating massive main-sequence stars \citep{Decressin_2007,Krause_2013, Yang_2013}, and massive binaries \citep{de_Mink_2009, 10.1093/mnras/stt1745} are considered the main polluters in GC systems. However, this interpretation remains unclear to date, and the nature of the polluters is uncertain \citep{Bastian_2018}.\\
It is worth noting that although large spectroscopic surveys such as the Apache Point Observatory Galactic Evolution Experiment (APOGEE; \citealt{Majewski_2017}) within the Sloan Digital Sky Survey-IV \citep{Blanton_2017} have made major contributions to the field, only a limited number of relatively metal-rich GCs have been studied in detail \citep{10.1093/mnras/stw3093,Johnson_2018,2020MNRAS.492.1641M}. This likely arises because most of the GCs in the metal-rich regime are concentrated in the Galactic bulge, where high interstellar extinction and stellar crowding complicate analysis. Significant efforts in this direction have been done by the bulge Cluster APOgee Survey (CAPOS) survey \citep{2021A&A...652A.157G}.\\

In this context, we focus our analysis on Patchick 126, a poorly studied GC candidate included in the compilation by \citet{2024A&A...687A.214G}. It was suggested to be a genuine, metal-rich GC located in the Galactic bulge. Section \ref{sec:stateofpat126} outlines the current knowledge of Patchick 126 and summarises the existing information on this star cluster. Section \ref{sec:spectroscopicanalysis} describes the membership selection criteria adopted to identify the Immersion GRating INfrared Spectrometer (IGRINS) targets. Section~\ref{sec:atmospheric_parameters} then presents, for the first time, the derivation of their atmospheric parameters, followed by the chemical abundance analysis described in Section \ref{Sec:comp_el_abu}. The resulting $\mathrm{[Fe/H]}$ and $\mathrm{[\alpha/Fe]}$ ratios are then used to independently constrain the cluster’s reddening, distance, age, and metallicity through the Hubble Space Telescope (HST) photometric analysis described in Section \ref{sec:phot_analysis}. Our main results and conclusions are summarised in Section \ref{sec:discussion_conclusion}. Finally, we also compare these with Galactic open and GCs in Appendix \ref{sec:discussions}.

\section{Current state of Patchick 126}
\label{sec:stateofpat126}
\begin{figure}
    \centering
    \includegraphics[width=1\linewidth]{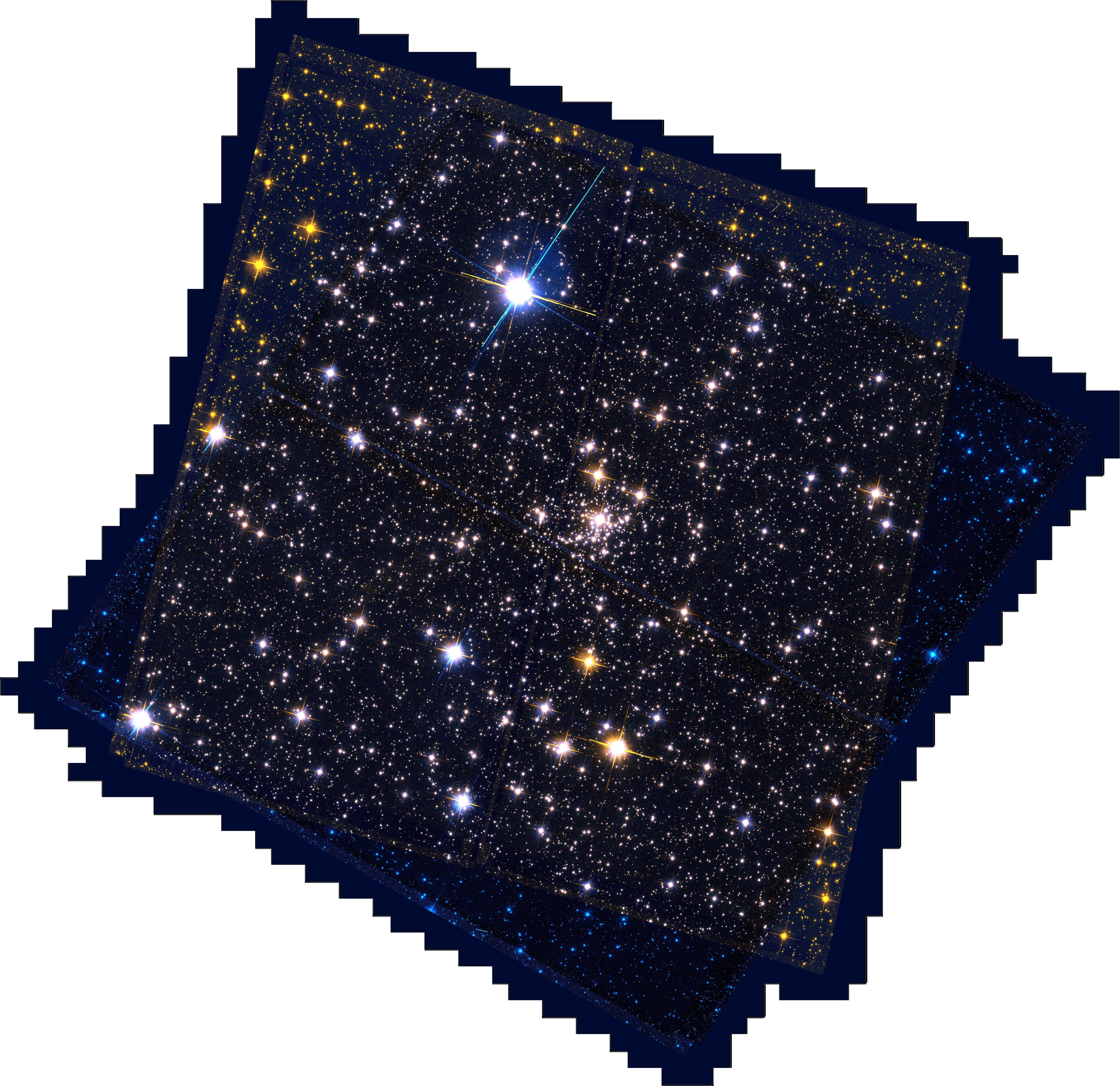}
    \caption{Colour-stacked image (red: F814W filter; blue: F606W filter; green: false colour from the average of the other two channels) of the central field of Patchick~126 covered by the \textit{Hubble} MGCS ACS/WFC data.}
    \label{Fig:HST_picture}
\end{figure}

Patchick 126 (Figure \ref{Fig:HST_picture}), located at RA = 17:05:38.6 and Dec = -47:20:32, is one of the least studied star clusters in the \cite{2024A&A...687A.214G} compilation. Initially discovered in the VVVX tile \texttt{e676}, it was first analysed in detail by \cite{2022A&A...659A.155G, 2023A&A...669A.136G}, who classified it as a Galactic GC with a low mass of$M\approx 2.4 \times 10^{3}\ M_{\odot}$. According to their photometric analysis, the cluster exhibits a well-defined RGB and a sparsely populated RC in its CMD (see Panel(a) of Figure \ref{Fig:selection}) -- features used to fit isochrone models, yielding an age $>8$ Gyr and a metallicity of [Fe/H] = $-0.7 \pm 0.3$. A complementary $\chi^{2}$ analysis provided an independent, somewhat younger age estimate of 5.6 Gyr. This discrepancy underscores the challenges of age determination, particularly without the main sequence turn-off at the CMD.

Furthermore, \cite{2023A&A...669A.136G} derived a mean radial velocity (RV) of $-121.8 \pm 0.3$ km s$^{-1}$ and computed the cluster’s orbital parameters using the proper motion values $\mu_{\alpha_{\ast}} = -4.96 \pm 0.40$ and $\mu_{\delta} = -6.80 \pm 0.40$ mas yr$^{-1}$. They found that Patchick 126 follows a moderately eccentric ($e \approx 0.35$), prograde orbit, plunging to a perigalactic distance of $r_{\mathrm{peri}} = 1.6$ kpc and reaching an apogalactic distance of $r_{\mathrm{apo}} = 4.2$ kpc, with a vertical excursion of 0.65 kpc above the Galactic plane, indicating an inner disc-like orbit. At the time of observation, the cluster was located within the MW bulge, at a heliocentric distance of $\sim 8$ kpc and a galactocentric distance of $\sim 2.8$ kpc.

The present work focuses on the combination of photometric data from \textit{HST} and spectroscopic information from IGRINS. For the first time, we are able to disentangle the true nature of this cluster, break the age-metallicity degeneracy, derive a precise age estimate, and trace its chemical evolution with unprecedented precision.

\section{IGRINS/Gemini-South observations}
\label{sec:spectroscopicanalysis}
\begin{figure}
    \centering
    \includegraphics[width=1\linewidth]{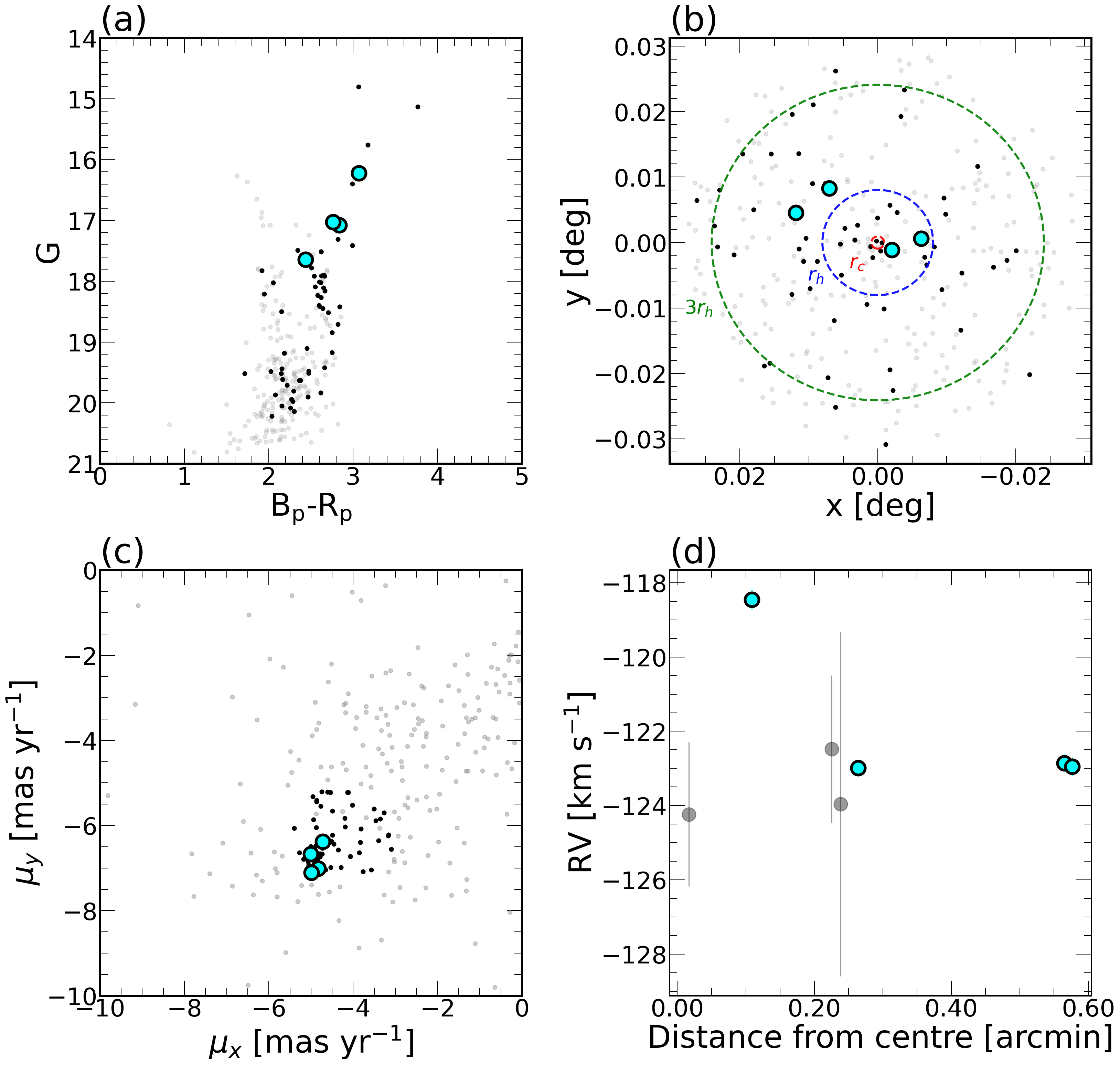}
    \caption{Targets selection. {\bf (a)} \textit{Gaia} DR3 CMD for field stars selected within $\sim3$ arcmin from the cluster centre (grey points) and for the high-probability proper-motion cluster members (black points). {\bf (b)} Positions of the two samples shown in the (x, y) coordinates. The x-axis has been inverted so that it corresponds to the usual inversion of right ascension. The red, blue, green circles draw the core radius $r_c\sim 0.056$ arcmin, half-light radius $r_h \sim 0.482$ arcmin, and 3$r_h$, respectively. {\bf (c)} Vector proper-motion diagram in (x, y) coordinates. {\bf (d)} RV vs distance from the cluster centre for the four stars listed in Table \ref{Table:RVigrins}, along with three additional stars with \textit{Gaia} RVs, listed in \cite{2023A&A...669A.136G}. In all panels, the cyan points represent the four giant stars with IGRINS data.}
    \label{Fig:selection}
\end{figure}

As detailed in \cite{2023A&A...669A.136G}, we observed four candidate RGB member stars with $K_{s} < 12$ in the direction of the GC candidate Patchick 126. As illustrated in Figure \ref{Fig:selection}, we selected the target stars according to three main criteria: \textit{(i)} their location along the RGB evolutionary sequence in the CMD (Panel (a)); \textit{(ii)} their proximity to the cluster centre, using core, half-light, and tidal radius from the \cite{2019MNRAS.482.5138B} catalogue\footnote{\url{https://people.smp.uq.edu.au/HolgerBaumgardt/globular/parameter.html}}, equal to $r_c=0.056$ arcmin, $r_h = 0.482$ arcmin, and $r_t=5.55$ arcmin, at a distance $\approx 8$ kpc (these stars fall within the cluster's tidal radius; see Panel (b)); and \textit{(iii)} their {\it Gaia} proper motions (Panel (c)). In particular, \citet{2022A&A...659A.155G} computed membership probabilities based on proper motions, and we retained only stars with probabilities higher than 80\%. Finally, Panel (d) displays the RVs from \citet{2023A&A...669A.136G} as a function of distance from the cluster centre, confirming that the selected stars share a coherent kinematic pattern.\\
In Figure \ref{Fig:selection}, we transformed the (RA, Dec) coordinates into Cartesian (x, y) coordinates. To define the orthographic projection of both the positional and proper motion components, we adopted Equation (2) from \cite{2018A&A...616A..12G}, as reported below:

\begin{align*}
    x &= \cos\delta\sin(\alpha - \alpha_{C}),\\
    y &= \sin\delta\cos\delta_{C} - \cos\delta\sin\delta_{C}\cos(\alpha - \alpha_{C}),\\
    \mu_{x} &= \mu_{\alpha_{\ast}}\cos(\alpha - \alpha_{C}) - \mu_{\delta}\sin\delta\sin(\alpha-\alpha_{C}), \\
    \mu_{y} &= \mu_{\alpha_{\ast}}\sin\delta_{C}\sin(\alpha - \alpha_{C}) \\ 
            &  + \mu_{\delta}(\cos\delta\cos\delta_{C} + \sin\delta\sin\delta_{C}\cos(\alpha - \alpha_{C})),
\end{align*}
where $\alpha_{C}$ and $\delta_{C}$ are the Patchick 126 centre coordinates.\\

Observations were carried out in queue mode as part of programme \texttt{GS-2022A-Q-238} (PI: E. R. Garro), using the then-visiting high-resolution spectrograph IGRINS (R $\sim$ 45,000), mounted on the 8.1 m Gemini-South telescope at Cerro Pachón, Chile. IGRINS is a cross-dispersed spectrograph with two arms that simultaneously cover the H and K bands, offering a continuous wavelength range from 1.45 to 2.45 $\mu$m \citep{2018SPIE10702E..0QM}. Each target was observed using an \texttt{ABBA} nodding sequence along the slit, along with a nearby \texttt{A0V} telluric standard star for atmospheric correction. All details of the reduction procedure are provided in \citet{2023A&A...669A.136G}.  A summary of the observational setup for each target is provided in Table~\ref{Table:igrinsprog}, and their kinematic properties are listed in Table~\ref{Table:RVigrins}.

\begin{table*}[htpb]
\centering 
\renewcommand{\arraystretch}{1.2}
\caption{Program star parameters and observations with the IGRINS spectrograph for Patchick 126, taken from \cite{2023A&A...669A.136G}.}
\begin{tabular}{lcccccccccccc}
\hline\hline
Nstar\tablefootmark{(a)} & ID\tablefootmark{(b)}      & RA     &  Dec    &   K\tablefootmark{(b)} & Date\tablefootmark{(c)}         & Exposure & S/N$_{H}$ & S/N$_{K}$\\
       &                  &  [hh:mm:ss]&  [dd:mm:ss] &    [mag]  &   (2022)       &  (s)    &   (pixel$^{-1}$) & (pixel$^{-1}$) \\
\hline
Star 1 &	7053569-4720154 &	17:05:35.68 &-47:20:15.5 & 12.310  & April 2 & 341.75 & 157.85 & 157.11 \\
Star 2 &	7053688-4720019 &	17:05:36.88 &-47:20:02.2 & 10.991  & April 2 & 101.5   & 166.02 & 141.05\\
Star 3 &	7054018-4720297 &	17:05:40.14 &-47:20:29.8 & 11.917 & April 23 & 238.0 & 156.15 & 162.83\\
Star 4 &	7053909-4720359 &	17:05:39.10 &-47:20:36.1 & 12.435 & April 20 & 383.5 &128.64 & 128.78\\
\hline\hline
\end{tabular}
\tablefoot{
         \tablefoottext{a}{We adopt this notation for ease of reference throughout the analysis.}
         \tablefoottext{b}{From the 2MASS \citep{2006AJ....131.1163S}.}
        \tablefoottext{c}{Date of the IGRINS observations.}
        }
\label{Table:igrinsprog}
\end{table*}

\begin{table*}[htpb]
\centering 
\renewcommand{\arraystretch}{1.2}
\caption{Kinematical properties of the Patchick~126 GCs observed with the IGRINS spectrograph, taken from \cite{2023A&A...669A.136G}.}
\begin{tabular}{lccccccccc}
\hline\hline
Nstar & $\mu_{\alpha_{\ast}}$\tablefootmark{(a)}  & $ \mu_{\delta}$\tablefootmark{(a)}   &  $\mu_{\alpha_{\ast}}$\tablefootmark{(b)}  &  $ \mu_{\delta}$\tablefootmark{(b)} & $V_{helio}$ & $RV_{H}$& $\sigma_{RV_{H}}$ & $RV_{K}$ &$\sigma_{RV_{K}}$\\
&[mas/yr]& [mas/yr]& [mas/yr]& [mas/yr]		& [km s$^{-1}$]  &[km s$^{-1}$] &[km s$^{-1}$] &[km s$^{-1}$] &[km s$^{-1}$]\\
\hline
Star 1 &-5.298 & -6.875 &-4.807 & -7.026  & 20.91 & -122.86 & 0.24 &  -122.67 & 0.25\\
Star 2 & -5.123 & -6.359 &-4.711 & -6.388 & 20.83 & -122.95 & 0.25 &  -122.85 & 0.25\\
Star 3 & -4.626 & -6.882 &-5.018 & -6.664  & 19.60 & -122.99 & 0.22 &  -122.96 & 0.25\\
Star 4 & -4.384 & -6.369 &-4.995 & -7.101  & 20.74 & -118.45 & 0.27 &  -118.85  &0.25\\
\hline\hline
\end{tabular}
\tablefoot{
        \tablefoottext{a}{From the VVVX \citep{2024A&A...689A.148S}.}
        \tablefoottext{b}{From the {\it Gaia} data release 3 \citep{2023A&A...674A...1G}.}}
\label{Table:RVigrins}
\end{table*}

\section{Atmospheric parameters}
\label{sec:atmospheric_parameters}
To derive precise chemical abundances from the spectra, it is essential to first determine the following atmospheric parameters: effective temperature ($T_{\rm eff}$), surface gravity ($\log g$), and microturbulent velocity ($\xi_{t}$). Since these stars have not been previously observed, no comparison with values from the literature is possible.\\

We derived the effective temperatures ($T_{\rm eff}$) using the line depth ratios (LDRs) technique \citep{2023ApJ...949...86A}. This technique is based on the ratio of central depths of carefully selected absorption lines that exhibit markedly different sensitivities to $T_{\rm eff}$. The underlying principle is that spectral lines arising from transitions with low and high excitation potentials (EP) respond differently to changes in temperature. Lines with low EP are highly sensitive to $T_{\rm eff}$ variations, while those with high EP are comparatively less affected. As a result, the ratio of their depths serves as a robust temperature diagnostic. Although other stellar parameters, such as metallicity, surface gravity, and microturbulence also influence line strengths, the method specifically targets line pairs in which temperature sensitivity dominates. Additionally, using ratios helps mitigate systematic effects from atmospheric parameters that similarly influence both lines, thereby enhancing the method's reliability. Specifically, we followed the methodology outlined by \citet{2023ApJ...949...86A} for both the H and K bands, and by \citet{10.1093/mnras/stz237} for the H band. In practice, we first applied the LDR-$T_{\rm eff}$ calibrations from each reference independently, then adopted the average of the two results as the final temperature estimate for each target. We used mainly Fe~I, Ti~I, and Ti~II lines available within the relevant wavelength ranges, since these are more sensitive to changes in $T_{\rm eff}$. By comparison, we also estimated the effective temperatures from Gaia photometry using the relations by \cite{2021A&A...653A..90M}. We found very good agreement within the uncertainties for Star 1 ($T_{\rm eff}=4397$ K) and Star 3 ($T_{\rm eff}=4425$ K). For Star 2 ($T_{\rm eff}=3993$ K), the two methods differ by 114 K, which is marginally larger than the adopted uncertainty (listed in Table \ref{Table:atm_par}) but still statistically compatible within the expected errors within $1.2\sigma$. For Star~4, we obtained a slightly higher value ($T_{\rm eff}=4934$ K). However, we preferred to adopt the spectroscopic determinations, as stellar absorption lines are unaffected by interstellar extinction and our measurements are therefore independent of the photometric modelling assumptions.\\

In the absence of reliable Fe II lines in the IGRINS near-infrared spectral range, we derived the surface gravity of our giant stars by imposing an ionisation balance between neutral and ionised titanium lines (Ti~I and Ti~II). 
The underlying principle relies on the Saha equation, which describes the ionisation equilibrium as a function of temperature and electron pressure, the latter being closely tied to surface gravity. At a given $T_{\rm eff}$, a lower gravity implies lower electron pressure and therefore a higher ionisation fraction, increasing Ti II abundances relative to Ti I.
We adopted effective temperatures from the LDR method and determined the microturbulence velocity by minimising the trend between Fe I abundances, calculated with the equivalent width (EW) method using the \texttt{pyEW} code\footnote{\url{https://github.com/madamow/pyEW}}. The surface gravity was then adjusted iteratively until the abundances inferred from the Ti I and Ti II lines converged within the observational uncertainties (⟨[TiI/H]⟩=⟨[TiII/H]⟩). Although non-local thermodynamic equilibrium (NLTE) effects can influence the abundances of Ti I in cool giants, studies have shown that these effects are generally moderate in the metallicity range explored here and weaker than for Fe I (e.g. \citealt{Bergemann_2012}). 
Both $\log g$ and $\xi_{t}$ were derived using the Python wrapper for MOOG code \citep{1973ApJ...184..839S,2012ascl.soft02009S}, named \texttt{pymoogi}\footnote{\url{https://github.com/madamow/pymoogi}}.\\
Nonetheless, using the derived values of $T_{\mathrm{eff}}$ and $\log g$, we independently determined $\xi_{t}$ with the Brussels Automatic Stellar Parameter code (\texttt{BACCHUS}; \citealt{2016ascl.soft05004M}). The results are consistent with those obtained from the previous method, within the errors. The adopted atmospheric parameters are listed in Table \ref{Table:atm_par}.\\

Finally, we also determined the abundance ratios by first adjusting the convolution parameter with a Gaussian broadening of $-9.05$ km s$^{-1}$ in order to account for the combination of stellar and instrumental effects. This convolution parameter is necessary to match observed and synthetic spectra. For this task, the broadening parameter was verified by comparing the abundances from strong, clear, and unsaturated Fe I lines -- derived from EW and minimum $\chi^2$ measurements -- and confirming good agreement. After the convolution parameter was fixed, we performed one iteration over Fe to obtain the metallicity ([Fe/H]) and three to four iterations over C, N, and O to fit the molecular features (see Section \ref{sec:CNO}).

\begin{table}[htpb]
\centering 
\renewcommand{\arraystretch}{1.2}
\caption{Atmospheric parameters for the four stars belonging to Patchick 126.}
\begin{tabular}{lccc}
\hline\hline
Nstar &  $T_{\rm eff}$ & $\log g$ & $\xi_{t}$\\
       &  [K]&  [cgs] &    [km s$^{-1}$]  \\
\hline
Star 1 & $4397\pm 100$ & $1.60\pm0.3$ & $2.12\pm0.1$	\\
Star 2 & $4107\pm 100$ & $1.00\pm0.3$ & $2.46\pm0.1$	\\
Star 3 & $4425\pm 100$ & $1.50\pm0.3$ & $2.33\pm0.1$	\\
Star 4 & $4594\pm 100$ & $2.56\pm0.3$ & $1.28\pm0.1$	\\
\hline\hline
\end{tabular}
\label{Table:atm_par}
\end{table}

\section{Elemental abundances}
\label{Sec:comp_el_abu}

We used the \texttt{BACCHUS} code \citep{2016ascl.soft05004M} to determine the elemental abundances of sixteen chemical species -- including the $\alpha$- (O, Mg, Si, Ca, and Ti), light- (C, N), odd-Z (Na, Al), iron-peak (Fe, Co, Ni, Cr, Mn, and V), and \textit{s}-process (Ce) elements -- in the \textit{H} and \textit{K} bands, for the first time for four stars in the newly discovered bulge cluster Patchick~126.\\
Abundances ratios were derived under the assumption of local thermodynamic equilibrium (LTE), using \texttt{BACCHUS} in combination with \texttt{MARCS} model atmospheres \citep{2008A&A...486..951G}. These models include convection via standard mixing-length theory and employ detailed opacity sampling with over $\sim 108,000$ wavelength points to account for line blanketing. After determining the atmospheric parameters, as described in Section~\ref{sec:atmospheric_parameters}, we derived the metallicity from Fe~I lines. This initial metallicity estimate, along with the atmospheric parameters, was then held fixed to compute the abundances of individual chemical species, as described below.

We computed the chemical abundances for each elemental species using the following five-step approach both for the H and K bands (Table \ref{Table:abundances_HKBAND}):
(a) Spectral synthesis was performed using the complete atomic line list from \citet{2021AJ....161..254S}, internally labelled as \texttt{linelist.20170418} (reflecting its creation date in YYYYMMDD format), for the H band. This line list was also used to estimate the local continuum level via a linear fit. In contrast, we used the \cite{Afşar_2018} and \cite{Guerço_2019} line lists for the K band.
(b) Cosmic rays and telluric lines were identified and rejected.
(c) The local signal-to-noise ratio (S/N) was estimated.
(d) A set of flux points contributing to each absorption line was automatically selected.
(e) Abundances were then derived by comparing the observed spectrum to a grid of convolved synthetic spectra with varying elemental abundances.
We subsequently applied four different diagnostic methods to determine the final abundances: (1) line-profile fitting, (2) core line-intensity comparison, (3) a global goodness-of-fit estimate, and (4) EW comparison. Each method yields a validation flag, and a decision tree was used to accept or reject each estimate, ultimately selecting the best-fit abundance. Among the available diagnostics, we adopted the $\chi^{2}$ diagnostic as the final abundance indicator due to its robustness. It is noteworthy that the differences between the four methods remain within the reported uncertainties.\\

The resulting elemental abundance ratios are listed in Table~\ref{Table:abundances_HKBAND}, which are scaled to the solar reference values of \citet{2006NuPhA.777....1A}. Uncertainties were estimated by individually varying each atmospheric parameter by $\Delta T_{\rm eff}=\pm 100$ K, $\Delta \log g=\pm 0.3$ cgs, and $\Delta \xi_{t} = \pm 0.1$ km s$^{-1}$, values commonly adopted in the literature (e.g. \citealt{2019A&A...627A.178F}). These values were chosen as they reflect the typical uncertainties in the atmospheric parameters for our sample. The reported uncertainties are therefore defined as the quadrature sum of the individual contributions from each parameter variation: 
\begin{equation*}
\sigma_{\mathrm{tot}}^{2} = \sigma_{T_{\mathrm{eff}}}^{2} + \sigma_{\log g}^{2} + \sigma_{\xi_{t}}^{2}.
\end{equation*}

It is important to note that for our analysis we used the mean values for the common elements derived from both the \textit{H} and \textit{K} bands: Fe, Ca, Mg, Si, Al, and Ti (see Table \ref{Table:abundances_HKBAND}). However, when comparing the abundances derived from the H- and K-band analyses, we found slight discrepancies that we were unable to fully attribute to a single cause. These differences may arise from uncertainties in atmospheric parameters, convolution choices, atomic data, or possible NLTE effects, as also suggested by \cite{2022A&A...666A..62L}.\\
Overall, Figure \ref{Fig:XFe_FeH_abu} qualitatively reveals that almost all of the chemical species examined so far in Patchick~126 stars exhibit a chemical enrichment roughly similar to that seen in other Galactic GCs, such as NGC~6553 \citep{2006A&A...460..269A,10.1093/mnras/stw3093,10.1093/mnras/stw2739,10.1093/mnras/stab712}, NGC~6388 \citep{2020MNRAS.492.1641M,Minelli_2021, 2023A&A...677A..73C}, NGC~6441 \citep{2006A&A...455..271G, 2020MNRAS.492.1641M}, and NGC~6838 (M71) \citep{2020MNRAS.492.1641M}.

\begin{table*}[htpb]
\centering
\renewcommand{\arraystretch}{1.3}
\caption{Patchick 126 chemical abundances. LTE chemical abundances for the four giant stars from IGRINS \textit{H-} and \textit{K-}band spectra, along with mean cluster values for the cluster are shown.}
\begin{tabular}{lcccccccc}
\hline
\hline
Species & Star 1 & Star 2 & Star 3 & Star 4 &  Average & $1\sigma$ & Std & Spread \\
\hline
\multicolumn{9}{c}{\textbf{\textit{H} band}}\\
\hline
$\mathrm{[C/Fe]}$  &  $0.03\pm0.10$ &  $0.07\pm0.14$ &  $0.10\pm0.11$ &  $0.41\pm0.08$ &  $0.15\pm0.05$ & 0.11 & 0.17 & 0.38 \\
$\mathrm{[N/Fe]}$  &  $0.29\pm0.19$ &  $0.42\pm0.15$ &  $0.19\pm0.18$ & $-0.23\pm0.28$ &  $0.17\pm0.10$ & 0.19 & 0.28 & 0.65 \\
$\mathrm{[O/Fe]}$  &  $0.26\pm0.16$ &  $0.36\pm0.16$ &  $0.21\pm0.15$ &  $0.13\pm0.14$ &  $0.24\pm0.07$ & 0.07 & 0.10 & 0.23 \\
$\mathrm{[Mg/Fe]}$ &  $0.24\pm0.14$ &  $0.24\pm0.06$ &  $0.30\pm0.12$ &  $0.24\pm0.17$ &  $0.25\pm0.05$ & 0.02 & 0.03 & 0.06 \\
$\mathrm{[Al/Fe]}$ &  $0.27\pm0.12$ &  $0.19\pm0.13$ &  $0.30\pm0.12$ &  $0.26\pm0.13$ &  $0.26\pm0.06$ & 0.03 & 0.05 & 0.11 \\
$\mathrm{[Si/Fe]}$ &  $0.10\pm0.06$ & $-0.02\pm0.07$ &  $0.09\pm0.06$ &  $0.35\pm0.07$ &  $0.13\pm0.03$ & 0.10 & 0.16 & 0.37 \\
$\mathrm{[Ca/Fe]}$ &  $0.12\pm0.10$ &  $0.16\pm0.11$ &  $0.22\pm0.10$ &  $0.18\pm0.12$ &  $0.17\pm0.05$ & 0.03 & 0.04 & 0.10 \\
$\mathrm{[Ti/Fe]}$ &  $0.20\pm0.18$ &  $0.23\pm0.13$ &  $0.18\pm0.18$ &  $0.16\pm0.17$ &  $0.20\pm0.08$ & 0.02 & 0.03 & 0.07 \\
$\mathrm{[V/Fe]}$  &  $0.26\pm0.10$ &  $0.39\pm0.11$ &  $0.18\pm0.16$ &  \dots &  $0.29\pm0.07$ & 0.07 & 0.11 & 0.21 \\
$\mathrm{[Cr/Fe]}$ &  $0.17\pm0.11$ &  $0.19\pm0.11$ &  $0.23\pm0.12$ &  $0.32\pm0.11$ &  $0.23\pm0.06$ & 0.05 & 0.07 & 0.15 \\
$\mathrm{[Mn/Fe]}$ &  $0.19\pm0.07$ &  $0.21\pm0.06$ &  $0.12\pm0.08$ &  $0.23\pm0.08$ &  $0.19\pm0.04$ & 0.03 & 0.05 & 0.11 \\
$\mathrm{[Fe/H]}$  & $-0.25\pm0.05$ & $-0.35\pm0.03$ & $-0.32\pm0.05$ & $-0.24\pm0.06$ & $-0.29\pm0.02$ & 0.05 & 0.05 & 0.11 \\
$\mathrm{[Co/Fe]}$ &  $0.07\pm0.10$ &  $0.09\pm0.07$ &  $0.05\pm0.09$ &  \dots &  $0.07\pm0.05$ & 0.01 & 0.02 & 0.04 \\
$\mathrm{[Ni/Fe]}$ &  $0.05\pm0.07$ &  $0.00\pm0.06$ &  $0.16\pm0.06$ &  $0.17\pm0.07$ &  $0.10\pm 0.03$ & 0.07 & 0.08 & 0.17 \\
$\mathrm{[Ce/Fe]}$ &  $0.03\pm0.16$ &  $0.02\pm0.12$ & $-0.06\pm0.15$ & $-0.04\pm0.14$ & $-0.01\pm0.07$ & 0.04 & 0.04 & 0.09 \\
\hline
\multicolumn{9}{c}{\textbf{\textit{K} band}}\\
\hline
$\mathrm{[Na/Fe]}$ &  $0.18\pm0.12$ &  $0.30\pm0.13$ &  $0.32\pm0.12$ &  $0.22\pm0.13$ &  $0.25\pm0.06$ & 0.06 & 0.07 & 0.14 \\
$\mathrm{[Mg/Fe]}$ &  $0.30\pm0.07$ &  $0.40\pm0.06$ &  $0.37\pm0.08$ &  \dots &  $0.36\pm0.04$ & 0.03 & 0.05 & 0.10 \\
$\mathrm{[Al/Fe]}$ &  $0.27\pm0.13$ &  $0.23\pm0.15$ &  $0.38\pm0.14$ &  $0.12\pm0.18$ &  $0.27\pm0.07$ & 0.08 & 0.11 & 0.26 \\
$\mathrm{[Si/Fe]}$ &  $0.09\pm0.07$ &  $0.07\pm0.09$ &  $0.08\pm0.09$ &  \dots &  $0.08\pm0.05$ & 0.01 & 0.01 & 0.02 \\
$\mathrm{[Ca/Fe]}$ &  $0.07\pm0.14$ &  $0.16\pm0.15$ &  $0.21\pm0.13$ &  $0.04\pm0.17$ &  $0.13\pm0.07$ & 0.07 & 0.08 & 0.17 \\
$\mathrm{[Ti/Fe]}$ &  $0.34\pm0.17$ &  $0.43\pm0.18$ &  $0.36\pm0.18$ &  $0.18\pm0.16$ &  $0.32\pm0.09$ & 0.07 & 0.11 & 0.25 \\
$\mathrm{[Fe/H]}$  & $-0.30\pm0.05$ & $-0.45\pm0.05$ & $-0.42\pm0.07$ & $-0.10\pm0.04$ & $-0.28\pm0.03$ & 0.12 & 0.16 & 0.35 \\
\hline
\hline
\end{tabular}
\tablefoot{The weighted average for each elemental abundance, $1\sigma$ error, standard deviation (Std), and spread are shown for the full sample. $1\sigma$ is defined as (84th percentile - 16th percentile)/2). All of the listed elemental abundances are scaled to the solar reference value from \cite{2006NuPhA.777....1A}.}
\label{Table:abundances_HKBAND}
\end{table*}

\begin{figure*}[htpb]
    \centering
    \includegraphics[width=0.7\linewidth]{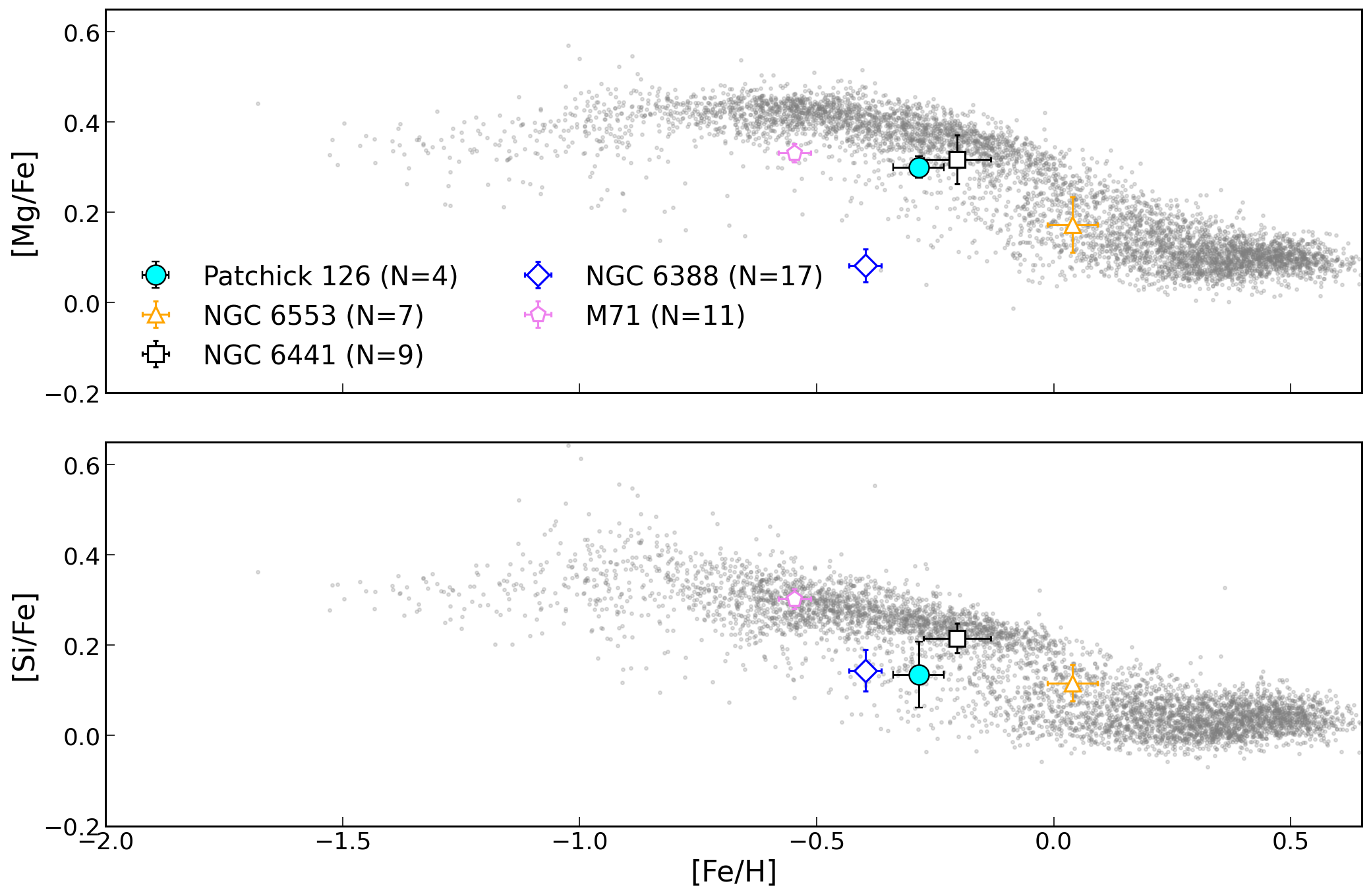}
    \caption{
    [Mg/Fe] and [Si/Fe] versus [Fe/H] diagrams for Patchick 126 (cyan points), NGC 6388 (open blue diamonds), NGC 6441 (open black squares), NGC 6838 (M71) (open violet pentagons), NGC 6553 (open orange triangles), and bulge field stars (grey points) at $R_{G}<3$ kpc. The APOGEE-2 bulge stars from \citet{Abdurrou2022} were selected using precise galactocentric distances from the {\ttfamily StarHorse} catalogue \citep{2023A&A...673A.155Q}. We rescaled the abundances derived by \citet{Abdurrou2022} using \texttt{ASPCAP} to match those obtained with \texttt{BACCHUS}, applying systematic offsets of 0.11, 0.07, and 0.05 for [Fe/H], [Mg/Fe], and [Si/Fe], respectively, as determined by \citet{2020A&A...643L...4F}.
    }
    \label{Fig:alphacomparison}
\end{figure*}

\begin{figure*}[htpb]
    \centering
    \includegraphics[width=1\linewidth]{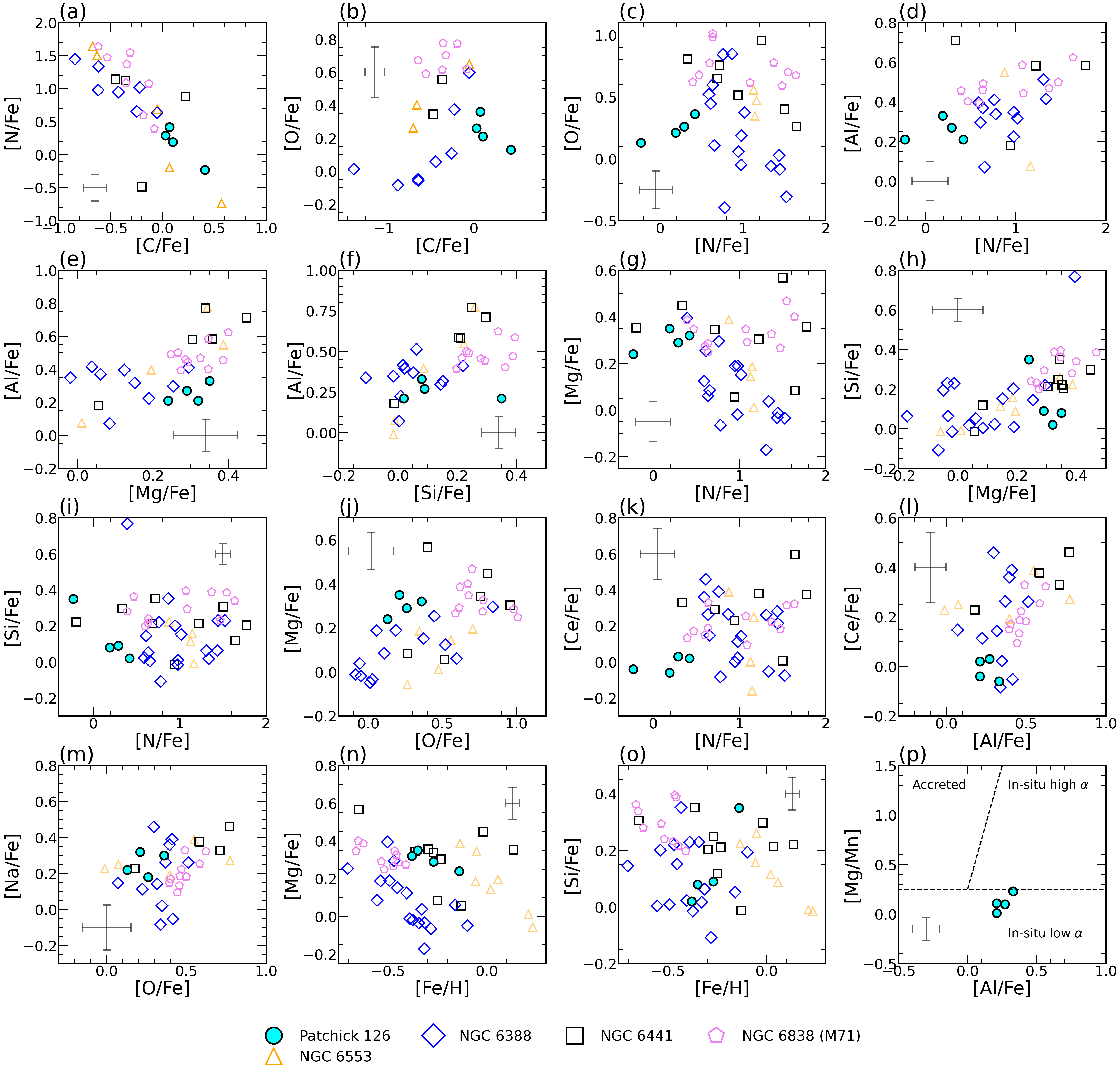}
    \caption{Combined $\alpha$-, light-, odd-Z, and s-process elements from our \texttt{BACCHUS} results listed in Table \ref{Table:abundances_HKBAND}. Notably, we used mean values for common elements in the \textit{H} and \textit{K} band. Panels (a)-(p): [C/Fe]-[N/Fe], [C/Fe]-[O/Fe], [N/Fe]-[O/Fe], [C/Fe]-[N/Al], [Mg/Fe]-[Al/Fe], [Si/Fe]-[Al/Fe], [N/Fe]-[Mg/Fe], [Mg/Fe]-[Si/Fe], [N/Fe]-[Si/Fe], [O/Fe]-[Mg/Fe], [N/Fe]-[Ce/Fe], [Al/Fe]-[Ce/Fe], [O/Fe]-[Na/Fe], [Fe/H]-[Mg/Fe], [Fe/H]-[Si/Fe], and [Al/Fe]-[Mg/Mn] distributions for Patchick 126 (cyan circles). We compare our results with other APOGEE-2 GCs examined by \cite{2020MNRAS.492.1641M}: NGC 6388 (open blue diamonds), NGC 6441 (open black squares), NGC 6838 (M71; open violet pentagons), and NGC 6553 (open orange triangles). The error bars mark the typical uncertainty ($\sigma_{tot}$) listed in Table \ref{Table:abundances_HKBAND}. The dashed black lines in Panel (p) define the criterion to separate in situ from accreted populations, similar to that defined by \cite{2021MNRAS.500.1385H}, \cite{2022A&A...663A.126F}, \cite{2023A&A...671A..45S}.}
    \label{Fig:XFe_FeH_abu}
\end{figure*}

\subsection{The $\alpha$-elements: O, Mg, Si, Ca and Ti}
\label{section:alphaelem}
Table \ref{Table:abundances_HKBAND} and Figure \ref{Fig:alphacomparison} show that Patchick 126 exhibits an $\alpha$-enhancement, with a mean value $\langle\mathrm{[O,Mg,Si,Ca,Ti/Fe]}\rangle = +0.19 \pm 0.02$ as derived from \textit{H-}band spectra and $\langle\mathrm{[Mg,Si,Ca,Ti/Fe]}\rangle = +0.18 \pm 0.06$ as derived from \textit{K-}band spectra. This result is consistent with expectations, as solar-metallicity clusters are not expected to show such enhancements (a typical value of $+0.3$ is usually observed in more massive, metal-poor GCs). Indeed, it remains consistent with $\alpha$-enhancement levels typically observed in thick-disc or inner-Galaxy GCs at similar metallicities, such as NGC~6553 and NGC~6528 \citep{2006A&A...460..269A, 10.1093/mnras/stab712, 2018A&A...620A..96M}. Given that the Galactic bulge extends to $\sim$3.5 kpc \citep{2016PASA...33...25Z, 2016ARA&A..54..529B,2018ARA&A..56..223B} and that Patchick~126 lies at a galactocentric distance of $\sim$2.8~kpc, these chemical properties further support its classification as an inner cluster.

\subsection{Derivation of C, N, and O abundances}
\label{sec:CNO}
The abundance ratios of [C/Fe], [N/Fe], and [O/Fe] have been derived from the molecular features $^{16}$OH, $^{12}$C$^{16}$O, and $^{12}$C$^{14}$N, following the same methodology described by \cite{2022A&A...657A..84F}. An iterative approach was adopted, starting with the determination of the oxygen abundances of the $^{16}$OH lines. Given that several $^{16}$OH lines are affected by telluric contamination, we selected a set of clean and well-defined lines individually for each star. Once the oxygen abundance was fixed, we derived carbon abundances of $^{12}$C$^{16}$O. For each star, we again identified the most suitable lines, noting that in the hotter target the $^{12}$C$^{16}$O features appeared weaker and more affected by blending, which made carbon determination more challenging. Nevertheless, we successfully obtained carbon abundances for all four stars. Finally, with both carbon and oxygen abundances fixed, nitrogen abundances were derived from $^{12}$C$^{14}$N molecular bands in the \textit{H} band. This sequence of steps was repeated iteratively until the changes in abundances between successive iterations became negligible, indicating convergence.\\

We found mean abundance ratios of $\langle\mathrm{[N/Fe]}\rangle = +0.17 \pm 0.10$, $\langle\mathrm{[C/Fe]}\rangle = +0.15 \pm 0.05$, and $\langle\mathrm{[O/Fe]}\rangle = +0.24 \pm 0.07$. Moreover, they appear to exhibit significant star-to-star variations, particularly in $\mathrm{[C/Fe]}$ and $\mathrm{[N/Fe]}$. We also found a modest but significant internal spread in carbon (observed $\sigma=0.17$ and intrinsic $\sigma=0.14$)\footnote{The observed $\sigma$ includes both real star-to-star differences and the contribution from measurement errors, whereas the intrinsic $\sigma$ is the de-convolved scatter after accounting for observational uncertainties.} Whereas, the largest intrinsic scatter is seen for nitrogen (observed $\sigma=0.28$ and intrinsic $\sigma=0.19$), which clearly varies from star to star.

The $\mathrm{[N/Fe]}$ abundance ratios show a broad spread, with the three stars showing N-enhancement $\mathrm{[N/Fe]}\geqslant 0.2$, whereas Star 4 is N-poor (0.6 lower than the mean value of N abundance; see Figure \ref{Fig:XFe_FeH_abu}). Despite the relatively large uncertainties, the amplitude of the variation exceeds the measurement errors, indicating an abundance dispersion. Finally, the $\mathrm{[O/Fe]}$ ratios, while showing a milder range (0.13 to 0.36), are consistent with $\alpha$-enhanced populations (see Section \ref{section:alphaelem}).

Figure \ref{Fig:XFe_FeH_abu}, panel (a), clearly reveals an apparent C--N anti-correlation, indicating the prevalence of the multiple-population phenomenon in Patchick~126 and supporting its classification as a genuine metal-rich GC. The same figure qualitatively compares the CNO behaviour of Patchick~126 with that of well-studied metal-rich GCs. We can see that Patchick~126 follows a similar behaviour to NGC~6553 and M~71, which also exhibit moderate C--N variations at similar metallicity. However, it is important to note that the C–N anti-correlation is based on only four stars. Stars 1, 2, and 3 show similar abundances, while Star 4 displays a lower nitrogen abundance. If we were to treat Star 4 — which exhibits slightly lower RVs but is located very close to the cluster's centre (i.e. Figure \ref{Fig:selection}, Panel (d)) — as an outlier, this anti-correlation would effectively disappear. Nevertheless, we prefer to retain Star 4 as a cluster member, as it satisfies all the membership criteria outlined in Section \ref{sec:spectroscopicanalysis}, although we interpret this result with due caution. Figure \ref{Fig:XFe_FeH_abu}, panel (c), reveals no O--N anti-correlation; oxygen abundances are stable, while nitrogen varies. This is similar to NGC 6553, NGC 6528, and M 71 and contrasts with the strong O--N anti-correlation visible in NGC 6441. Overall, the CNO abundances indicate that Patchick~126 behaves as other metal-rich clusters at similar metallicity.

\subsection{Odd-Z elements: Na, Al}
Sodium abundances were derived exclusively from \textit{K}-band spectra, using the lines at 22056.4, 22083.7, 23378.9, and 23379.1\,\AA. We obtained a mean ratio of $\langle\mathrm{[Na/Fe]}\rangle = +0.26 \pm 0.06$, indicating that stars in Patchick~126 are systematically Na-enhanced compared to field stars, which at $\mathrm{[Fe/H]}\sim -0.3$ typically show near-solar values ($\mathrm{[Na/Fe]} \approx 0.0$; \citealt{1987A&A...178..179G}). As shown in Figure~\ref{Fig:XFe_FeH_abu}, the [Na/Fe] ratios of Patchick~126 stars are consistent with those of other Galactic GCs at similar metallicity, supporting the view that the cluster exhibits the characteristic chemical pattern of bona fide GCs. However, with only four stars and observed dispersions of $\sigma\sim0.12{-}0.16$, we cannot establish the presence of an intrinsic Na--O spread or anti-correlation (Figure~\ref{Fig:XFe_FeH_abu}, Panel (m)). 

\begin{figure*}[htpb]
    \centering
    \includegraphics[width=0.7\linewidth]{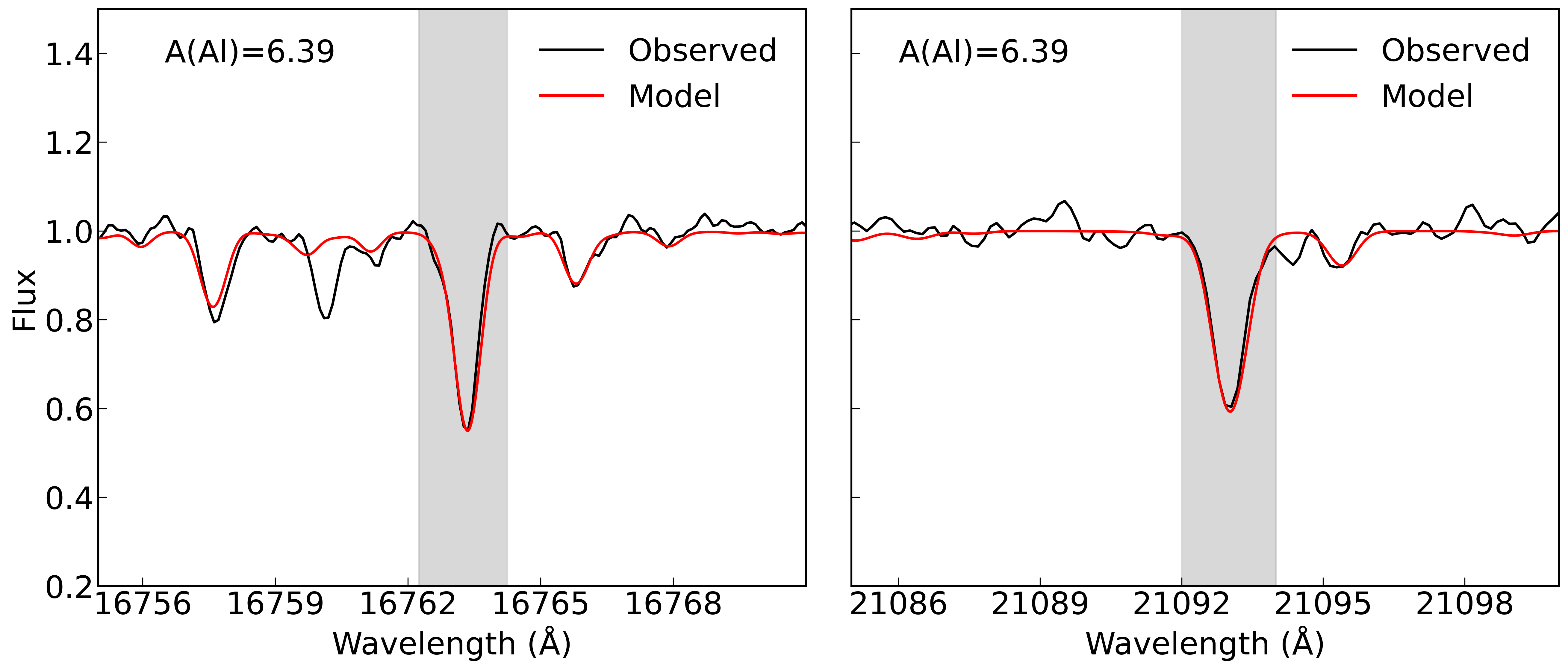}
    \caption{
Comparison of synthetic spectra (red lines) to the observed spectra (black line) for a Patchick 126 star, Star 1. Each panel shows the best-determined (grey-shaded region) [Al/Fe] abundance ratios for a representative line in the \textit{H} and \textit{K} bands, at 16763.4 $\AA$ and 21093.0 $\AA$, respectively.}
    \label{Fig:aluminiumline}
\end{figure*}

Another light odd-Z element measured in both the \textit{H} and \textit{K} bands is aluminium (Al; Table~\ref{Table:abundances_HKBAND}). At first, we found a discrepancy of $\sim0.3$ between the two bands, with higher abundances derived from the \textit{K} band. To investigate this, we tested our analysis on the benchmark star Arcturus, which closely matches our targets in metallicity and atmospheric parameters. We found that this discrepancy may be due to the log(gf) values sourced from the NIST database, which were adopted in \cite{Afşar_2018}. 
We therefore performed a reverse analysis to derive astrophysical $\log gf$ values for the lines at 21093.0 and 21163.8\,\AA. The revised values ($-0.11$ instead of $-0.40$ for 21093.0\,\AA, and $+0.29$ instead of $-0.01$ for 21163.8\,\AA) were validated against the Arcturus atlas spectrum, using the abundances and stellar parameters of \cite{Ramírez_2011}. The results show satisfactory agreement with the abundances reported by \cite{Ramírez_2011}. Applying these new $\log gf$ values to our sample stars yielded Al abundances consistent between the \textit{H}- and \textit{K}-band measurements (see Figure~\ref{Fig:aluminiumline}).

We find that Patchick~126 exhibits a mean aluminium enrichment of $\langle\mathrm{[Al/Fe]}\rangle = +0.26 \pm 0.06$, with no evidence of an intrinsic Al spread. This is in line with expectations for metal-rich clusters, where the lower H-burning temperatures in intermediate-mass AGB stars suppress the Mg--Al cycle, limiting Al production \citep[e.g.][]{2009A&A...499..835V}. Consistently, no Mg--Al anti-correlation is visible in our data (Figure~\ref{Fig:XFe_FeH_abu}, Panel (e)). Also, this distribution is another diagnostic to support the in situ versus ex situ origin. Comparing our results with those of \cite{2024MNRAS.528.3198B} (see their Figure 11), we can support an in situ origin for Patchick~126.  \\

To further test for light-element relations, we applied a Spearman rank correlation analysis. This non-parametric test is well suited to small, noisy GC datasets. As expected, no significant trends were found: Na versus O shows non-correlation ($\rho=0.00$, $p=1.00$)\footnote{How to interpret: $\rho \approx\ \pm 1$ indicates a strong positive and/or monotonic trend, while $\rho\approx\ 0$ means no monotonic relation. The p-value quantifies whether the observed $\rho$ occurs by chance.}, while Mg versus Al shows only a weak, statistically insignificant positive trend ($\rho=+0.63$, $p=0.37$). Overall, the four stars define a chemically homogeneous locus, with moderate Na and Al enhancements but no corresponding Mg or O depletion -- similar to patterns observed in other Galactic GCs.  \\

\subsection{The iron-peak elements: Fe, Co, Ni, Cr, Mn, and V}
We find a mean metallicity of $\langle\mathrm{[Fe/H]}\rangle = -0.29 \pm 0.02$ for Patchick~126. The observed dispersion agrees well with the measurement errors, so we find no evidence for a statistically significant metallicity spread. However, its high metallicity places Patchick~126 among metal-rich bulge clusters. 

We were able to determine five additional iron-peak elements: Co, Ni, Cr, Mn, and V. Although these elements have not been widely studied in bulge GCs, the available data are insufficient to provide a qualitative comparison with Patchick~126. However, reporting these elemental abundances provides valuable tracers of chemical evolution. They are primarily produced by Type Ia supernovae and massive stars \citep{1995ApJS..101..181W, 2003ApJ...592..404L}, which are key contributors to the enrichment of the interstellar medium; in contrast, Type II supernovae mainly synthesise $\alpha$-elements.

The Co abundance was derived using a single line at 16757.6 $\AA$, and the V abundance at 15924.8 $\AA$. No reliable Co and V measurements could be obtained for Star 4. Similarly, Cr abundance was based solely on the line at 15860.2 $\AA$. As a result, the values derived for Co, V, and Cr should be interpreted with caution.\\

As far as other iron-peak elements are concerned, Co, Ni, Cr, Mn, and V are overabundant relative to the Sun, with a mean of $+$0.17. The mean abundance ratios are $\langle\mathrm{[Co/Fe]}\rangle = +0.07 \pm 0.04$, $\langle\mathrm{[Cr/Fe]}\rangle = +0.23 \pm 0.05$, $\langle\mathrm{[Mn/Fe]}\rangle = +0.19 \pm 0.04$, $\langle\mathrm{[Ni/Fe]}\rangle = +0.10 \pm 0.04$, and $\langle\mathrm{[V/Fe]}\rangle = +0.29 \pm 0.07$. These values are broadly consistent with those observed in a few bulge GCs, such as NGC~6528 \citep{2018A&A...620A..96M} and NGC~6553 \citep{10.1093/mnras/stw2739}, as well as in bulge field stars \citep{2013A&A...559A...5B, 2014AJ....148...67J}. The uniformly over-solar iron-peak abundances observed in Patchick~126 could suggest an early and efficient chemical enrichment, likely dominated by Type Ia supernovae. 

More interestingly, Figure \ref{Fig:XFe_FeH_abu}, panel (p), reveals the distribution of Mg, Mn, and Al in the [Mg/Mn]--[Al/Fe] plane for the stars in Patchick~126. The observed low-[Mg/Mn] ratios place this cluster in the low-$\alpha$ regimen, classifying it as an in situ cluster. Patchick~126 chemically behaves as a typical GC, enriched in light elements by internal polluters but with iron-peak ratios likely shaped by core-collapse supernovae. Furthermore, comparing our [V/Fe] abundances with \cite{Minelli_2021} -- where iron peak elements (Sc, V, and Zn) probe the origin of known GCs -- we can conclude that Patchick~126 follows trends seen in MW field stars, GCs NGC 5927, and NGC 6496, supporting an in situ origin.

\subsection{The \textit{s}-process element: Ce}

We derived [Ce/Fe] abundance ratios from two lines: 15784.8 $\AA$ and 16595.2 $\AA$. In Patchick 126, the [Ce/Fe] abundance ratios are nearly solar, with a mean value of $\langle\mathrm{[Ce/Fe]}\rangle = -0.01 \pm 0.07$, showing no significant star-to-star dispersion. We can confirm \textit{s}-process non-enhancement in all the four stars in Patchick~126, likely supporting the pollution of this cluster by massive AGB stars \citep{Ventura_2016}.

We also note that the chemical behaviour of cerium in Patchick~126 seems consistent with other old, metal-rich bulge GCs such as NGC~6553 and NGC~6528 \citep{2020MNRAS.492.1641M}. From Figure \ref{Fig:XFe_FeH_abu}, panels (k-l), we can see that Patchick~126 has solar to mildly sub-solar Ce. 

\section{The age of Patchick~126}
\label{sec:phot_analysis}

\begin{table}
\centering 
\renewcommand{\arraystretch}{1.2}
\caption{Best-fit values for Patchick 126 obtained with CARMA.}
\begin{tabular}{cccc}
\hline\hline
[M/H] & $E(B-V)$ & $(m-M)_{0}$ & Age \\

[dex] & [mag] & [mag] & [Gyr] \\
\hline \\
\vspace{0.2cm}
$-0.13^{+0.03}_{-0.02}$ & $1.08^{+0.01}_{-0.01}$ & $14.47^{+0.03}_{-0.04}$ & $11.9^{+0.3}_{-0.4}$ \\
\hline\hline
\end{tabular}
\tablefoot{The CMD fits and corner plots are shown in Figure \ref{Fig:CARMA_CMDs}. The up-to-date table with all the results from the CARMA project can be found at: \url{https://www.oas.inaf.it/en/research/m2-en/carma-en/}}
\label{Table:CARMA_ages}
\end{table}

\begin{figure*}[!th]
   \centering
   \begin{minipage}{0.45\textwidth}
        \centering
        \includegraphics[width=1.0\textwidth]{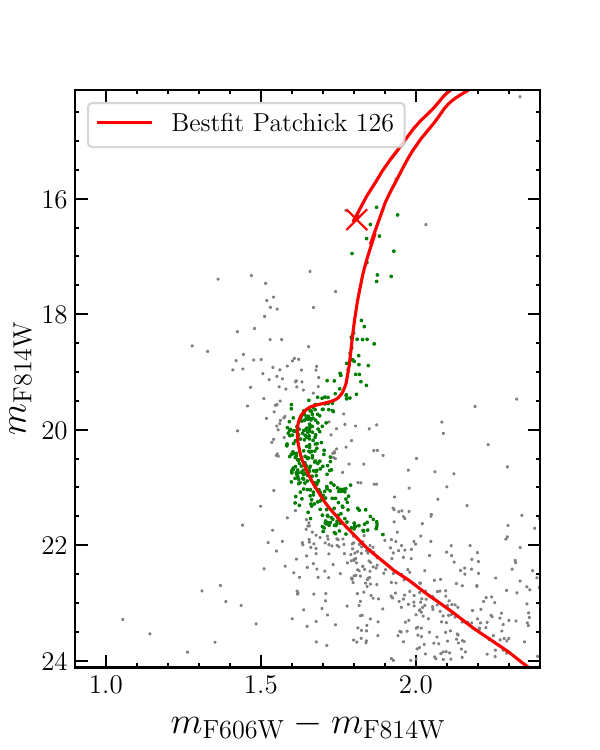}
        \includegraphics[width=1.0\textwidth]{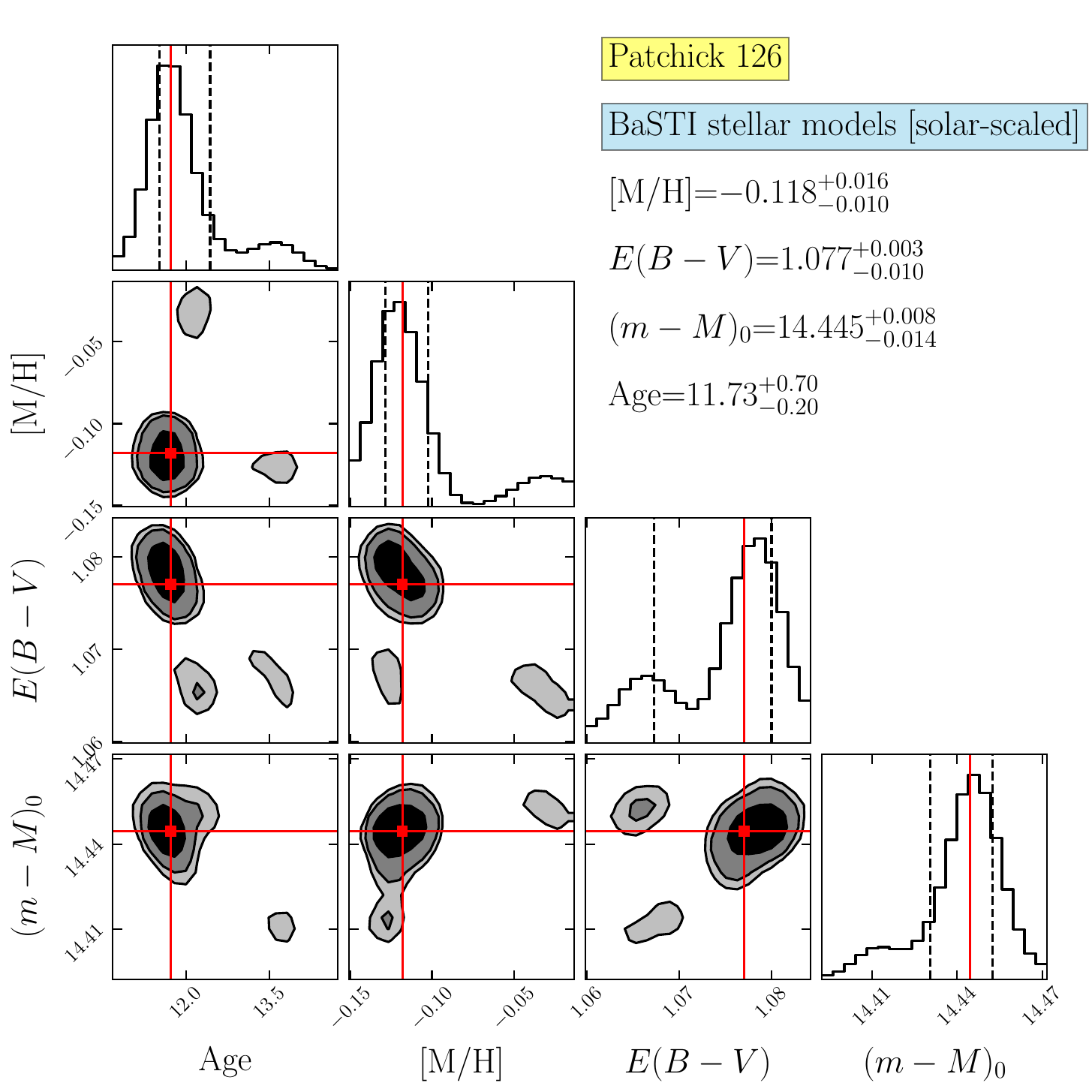}
   \end{minipage}\hfill
   \begin{minipage}{0.45\textwidth}
        \centering
        \includegraphics[width=1.0\textwidth]{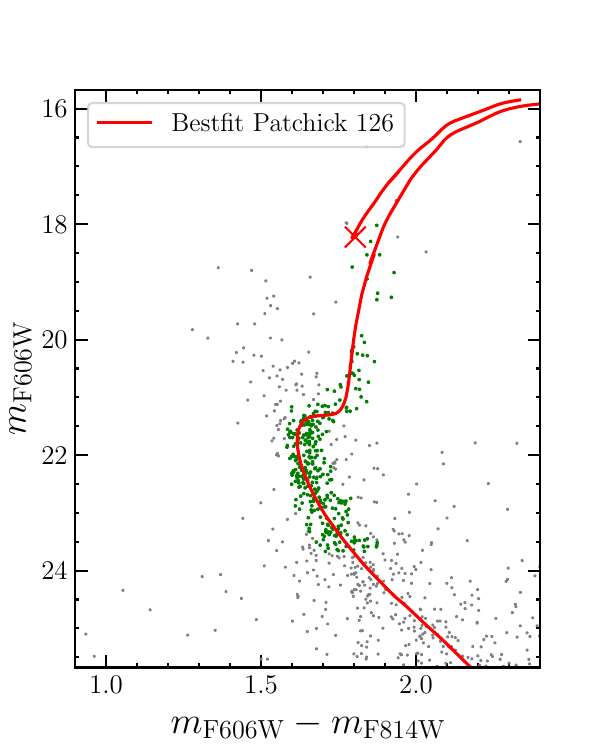}
        \includegraphics[width=1.0\textwidth]{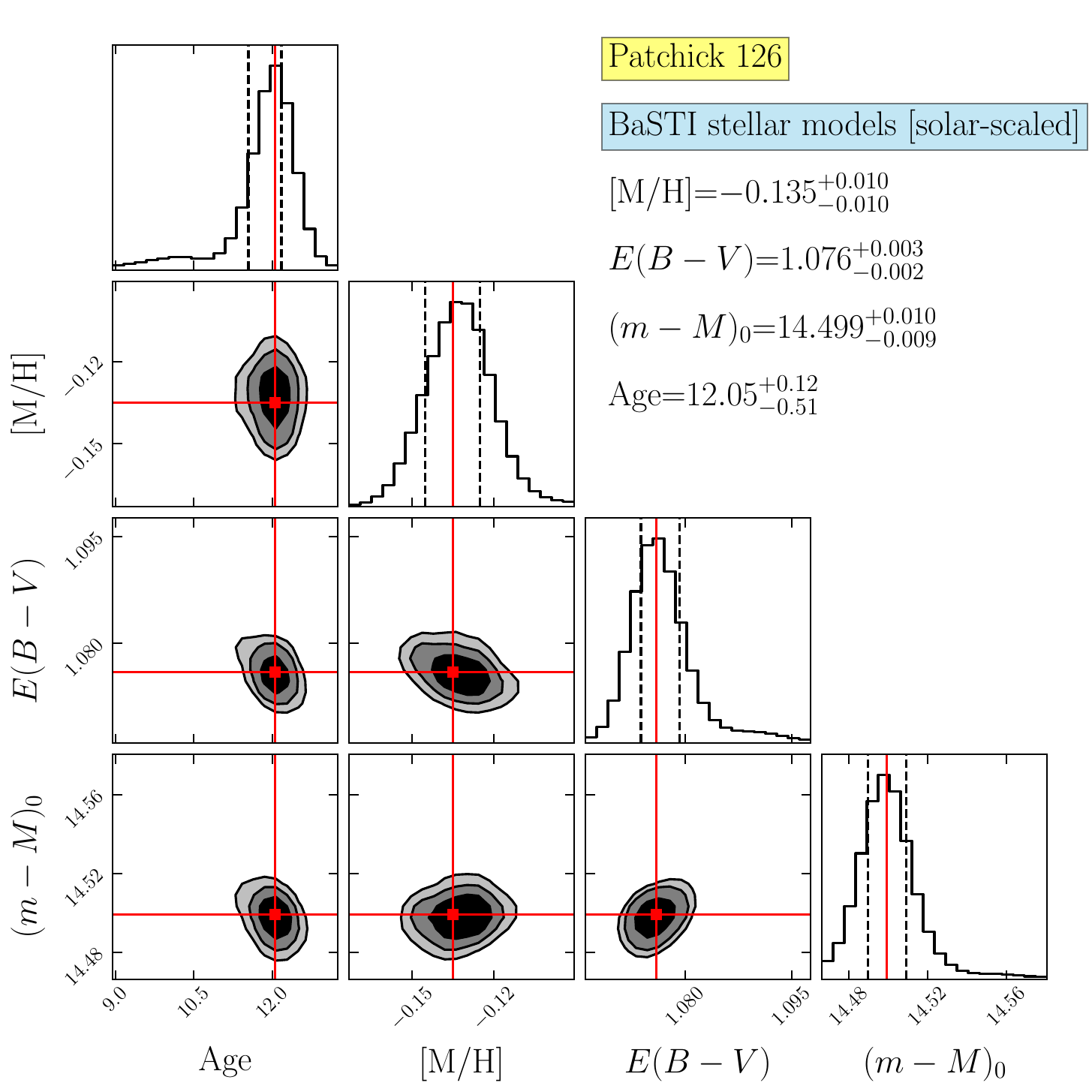}
   \end{minipage}
\caption{Results for Patchick 126. Top row: Best-fit model in the ($\it{m_{\mathrm{F814W}}}$, $\it{m_{\mathrm{F606W}}} - \it{m_{\mathrm{F814W}}}$) CMD (left) and in the ($\it{m_{\mathrm{F606W}}}$, $\it{m_{\mathrm{F606W}}} - \it{m_{\mathrm{F814W}}}$) CMD (right). Bottom row: Posterior distributions for the output parameters and the best-fit solution, quoted in the labels, in the ($\it{m_{\mathrm{F814W}}}$, $\it{m_{\mathrm{F606W}}} - \it{m_{\mathrm{F814W}}}$) CMD (left) and in the ($\it{m_{\mathrm{F606W}}}$, $\it{m_{\mathrm{F606W}}} - \it{m_{\mathrm{F814W}}}$) CMD (right).}
\label{Fig:CARMA_CMDs}%
\end{figure*}    
\begin{figure}
    \centering
    \includegraphics[width=1\linewidth]{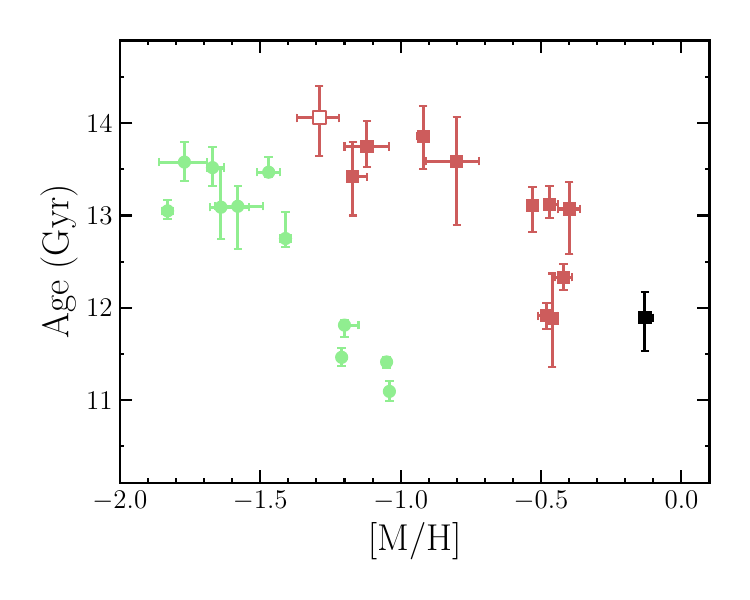}
    \caption{AMR derived within the CARMA project to date. Patchick 126 is shown as a black-filled square. In situ \citep[from~][]{massari2023,ceccarelli2025,massari2025} and \textit{Gaia}-Sausage-Enceladus \citep[from~][]{aguado-agelet2025} GCs are plotted as red- and green-filled symbols, respectively. 
    }
    \label{Fig:AMR}
\end{figure}

The age of Patchick 126 was derived from \textit{Hubble Space Telescope} photometric catalogues published as part of the MGCS survey \citep[][Libralato et al. in prep]{massari2025}, using the isochrone-fitting technique developed in the context of the Cluster Ages to Reconstruct the Milky Way Assembly (CARMA) project \citep{massari2023}. Following the standard procedures adopted within the CARMA collaboration, we employed {\it HST} photometry in the F606W and F814W bands to ensure homogeneity with previous CARMA age determinations \citep[see e.g.][]{aguado-agelet2025, ceccarelli2025, niederhofer25}. Since observations of parallel HST fields for Patchick~126 have failed, we could not apply any statistical decontamination to the cluster CMD. Moreover, no second-epoch {\it HST} observations exist; thus, kinematic membership is unfeasible as well. To select likely cluster members, we therefore restricted the sample to stars within the central region of the cluster, where the members dominate the most over field stars, adopting a radial cut of 25 arcsec corresponding to approximately one half-light radius\footnote{\url{https://people.smp.uq.edu.au/HolgerBaumgardt/globular/parameter.html}}. This sample of stars was corrected for differential reddening using the procedures described in \citep{milone2012}. We find a mild reddening variation across the field, up to $\Delta E(B-V)=0.04$ mag.

The method for our age determination is extensively described in \citet{massari2023}. Briefly, the isochrone fit provides a simultaneous solution for age, average reddening ($E(B-V)$), distance modulus ($(m-M)_{0}$), and global metallicity ([M/H]) within a Bayesian statistical framework. We adopted loose Gaussian priors on $E(B-V)$, $(m-M)_{0}$, and [M/H] centred on literature values of reddening and distance from \citet{2022A&A...659A.155G} and on the spectroscopic metallicity derived in this work (with $\sigma$ of 0.1, 0.1, 0.05, respectively). We used solar-scaled isochrones from the latest release of the BaSTI stellar evolution library \citep{hidalgo2018}, covering ages from 6 to 15 Gyr (with steps of 0.1 Gyr) and [M/H] values from -2.5 to 0.0 (with steps of 0.01). By working in terms of global metallicity and adopting solar-scaled models, we avoided assumptions about the $\alpha$-element abundances for Patchick 126, as solar-scaled and $\alpha$-enhanced models are equivalent at fixed [M/H] in the optical bands used here \citep{salaris1993}. Nonetheless, we used both the spectroscopic measurement of [Fe/H] and [$\alpha$/Fe] to determine [M/H], which we adopted as a prior to our fitting algorithm. This technique was applied separately to the two possible optical CMD combinations. In Figure \ref{Fig:CARMA_CMDs} we show the best-fit isochrones on top of the two CMDs in the top panels, while the posterior distributions for the free parameters are displayed in the bottom panels in the form of corner plots. For each parameter, the final result is the averaged value obtained from the two runs, while the uncertainties are calculated as the 16th and 84th percentiles of the posterior distribution of the two runs combined. The results are listed in Table \ref{Table:CARMA_ages}.

Following Eq. 1 of \citet{aguado-agelet2025}, [M/H] was converted into iron abundance assuming $\mathrm{[\alpha/Fe]} = 0.20$, derived by averaging mean abundances of O, Mg, Si, Ca, and Ti (as derived in Section \ref{section:alphaelem}). This yields $\mathrm{[Fe/H]} = -0.28$, in excellent agreement with the spectroscopic [Fe/H] measured in this work (see Section \ref{sec:spectroscopicanalysis}). The colour excess derived from the isochrone fitting confirms that Patchick 126 is a heavily reddened cluster with $E(B-V) = 1.08$ mag, although less extincted by $\sim 0.2$ mag compared to previous estimates \citep[e.g.][]{2022A&A...659A.155G}. Finally, we obtain an average heliocentric distance of $7.8^{+0.10}_{-0.14}$ kpc from the two runs, which is in line with the only other independent findings obtained from \textit{Gaia} and NIR 2MASS photometry \citep[][]{2022A&A...659A.155G}.  

Based on its mean age ($11.9$ Gyr) and metallicity ($\mathrm{[M/H] = -0.13}$), Patchick 126 is perfectly placed alongside the in situ branch of the MW GC age-metallicity relation (AMR) derived within the CARMA project \citep{massari2023,ceccarelli2025}, as shown in Figure \ref{Fig:AMR}. It is indeed $\sim 1.0$ more metal-rich than coeval GCs accreted from the \textit{Gaia}-Sausage-Enceladus dwarf galaxy \citep{belokurov2018,helmi2018,aguado-agelet2025}, suggesting a formation in a more massive galaxy capable of sustaining more efficient chemical enrichment, such as the MW. The in situ origin inferred from the age of Patchick 126 is consistent with the interpretation based on its orbital parameters, which we re-determined here using the distance inferred from isochrone fitting, positions, and proper motions from \cite{2021MNRAS.505.5978V}, and line-of-sight velocity from \cite{2023A&A...669A.136G}. By adopting the prescriptions on the MW potential as in \cite{2019A&A...630L...4M}, we find an orbital energy $E=-213659\pm 618$ km$^2$/s$^2$, a vertical angular momentum L$_z=539\pm13$ km/s~kpc, and a circularity of $0.89$. These values characterise Patchick~126 as moving on a disc-like, almost circular orbit, with a pericentre of $2.21\pm0.07$ kpc, an apocentre of $3.36\pm0.02$ kpc, an eccentricity of $0.2$, and a low-maximum vertical height of z$_{max}=0.57$ kpc, thus firmly verifying the criteria of GCs classified as in situ \citep[see][eDR3 edition]{2019A&A...630L...4M}\footnote{See the updated criteria in \cite{massarieDR3}}.  

\section{Conclusions}
\label{sec:discussion_conclusion}
This work presents the first comprehensive analysis of Patchick~126, combining high-resolution spectroscopy from IGRINS with deep \textit{HST} photometry. We derived elemental abundances for four red giant members in both the \textit{H} and \textit{K} bands (Table \ref{Table:abundances_HKBAND}), finding excellent agreement between the two datasets. Mean abundances for Fe, Ca, Mg, Si, Ti, and Al were adopted from this consistency. Despite the limited sample size and considering that all the target stars are cluster members, our results indicate that Patchick 126 is a metal-rich cluster with $\langle\mathrm{[Fe/H]}\rangle = -0.30 \pm 0.03$. The $\alpha$-elements (O, Mg, Si, Ca, and Ti) display a modest enhancement of $\sim 0.2$, in line with expectations for clusters at this metallicity.
We also constructed diagnostic anti-correlation diagrams, a hallmark of GCs. As discussed in Section \ref{sec:CNO}, a potential N--C anti-correlation is observed, which may thus suggest a GC nature for Patchick 126. However, if Star 4 (the only target with discrepant RV and chemical abundances) were excluded, this anti-correlation might disappear, weakening the case for the GC nature theory in favor of an old OC one -- even though no OC with such old age is known. Additionally, no evidence of other anti-correlations, such as Mg--Al and Na--O, is found for Patchick 126. This absence may result from the small sample size or could represent an intrinsic property of Patchick 126. Comparisons with well-studied Galactic GCs at similar metallicities suggest that strong anti-correlations are not expected -- and are often absent altogether -- in metal-rich or low-mass systems \citep{Pancino_2017}. 
Finally, we derived iron-peak elemental abundances, which closely resemble those of other metal-rich Galactic GCs such as NGC 6553 and NGC 6528. This similarity indicates that Patchick 126 follows the chemical trends of genuine GCs: enriched in light elements through internal polluters, but with iron-peak ratios primarily shaped by core-collapse supernovae. These conclusions are further supported by the behaviour of Ce, tracing the s-process contribution. \\
Patchick 126 is one of the poorly studied clusters included in the \textit{HST} MGCS survey. For the first time, we determined its precise age. By adopting the isochrone-fitting technique developed within the CARMA project, we derive an age of 11.9 Gyr and a global metallicity of [M/H] = $-0.13_{-0.02}^{+0.03}$ (equivalent to [Fe/H] = $-0.28$, assuming [$\alpha$/Fe] = +0.19 from the \citealt{aguado-agelet2025} relation) -- in excellent agreement with the spectroscopic analysis.\\ 
In conclusion, considering all the available evidence, we can firmly conclude that Patchick 126 is an in situ, low-mass, metal-rich, old system likely of GC-origin. Additional spectroscopic data will be essential to understand its true nature.


\begin{acknowledgements}
We thank Dr. Anna Barbara de Andrade Queiroz for sharing StarHorse data with us.\\

This work used the Immersion Grating Infrared Spectrometer (IGRINS) that was developed under a collaboration between the University of Texas at Austin and the Korea Astronomy and Space Science Institute (KASI) with the financial support of the US National Science Foundation 27 under grants AST-1229522 and AST-1702267, of the University of Texas at Austin, and of the Korean GMT Project of KASI.

E.R.G. gratefully acknowledges the ESO Fellowship program. GF-T gratefully acknowledges the grants support provided by ANID Fondecyt Iniciaci on No. 11220340, ANID Fondecyt Postdoc No. 3230001 and CAS-ANID/CAS220009 (sponsoring researcher in both), from the Joint Committee ESO-Government of Chile under the agreement 2021 ORP 023/2021 and 2023 ORP 062/2023.
BD acknowledges support from ANID Basal project FB210003. D.M. is supported by ANID Fondecyt Regular grant No. 1220724, and by the ANID BASAL Center for Astrophysics and Associated Tecnologies (CATA) through ANID grants ACE210002 and FB210003.
\end{acknowledgements}

\bibliographystyle{aa.bst}
\bibliography{biblio_spec}

\appendix
\section{Comparison with Galactic open and globular clusters}
\label{sec:discussions}
\begin{figure*}
    \centering
    \includegraphics[width=1.0\linewidth]{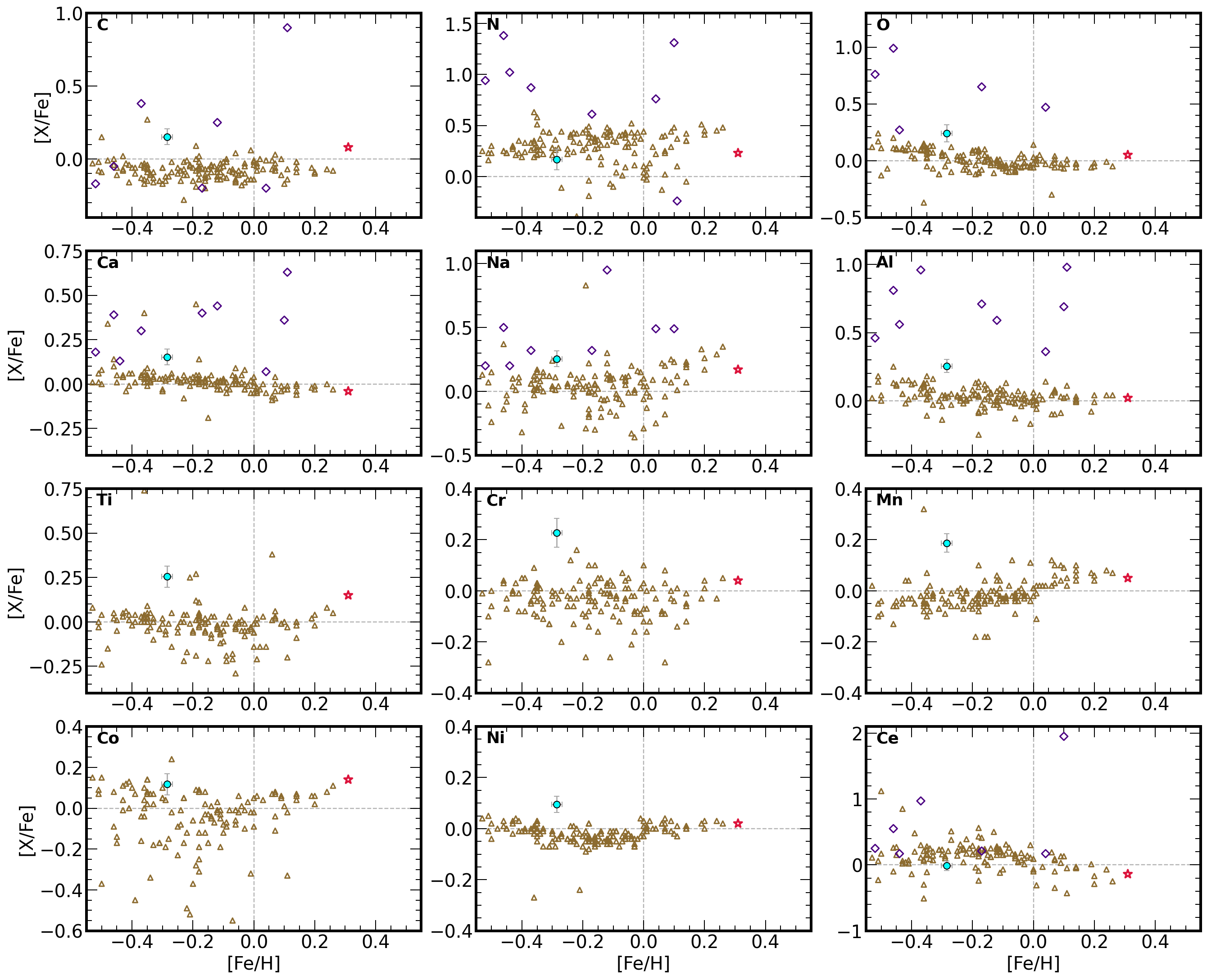}
    \caption{Comparison with OCs and GCs samples. We show the mean cluster abundance ratios as a function of [Fe/H], with the main aim to compare the chemistry of Patchick 126 (cyan circle) with that of old OCs from the \cite{Myers_2022} compilation (brown triangles) and with GCs from \cite{2020MNRAS.492.1641M} (purple diamonds) at the same metallicity range. The oldest known OC, NGC 6791, is highlighted with a red star.}
    \label{Fig:comparisonOCGCs}
\end{figure*}

Based on our analysis, Patchick 126 emerges as an intriguing system whose properties might place it at the boundary between globular and open cluster classifications. To contextualise its nature, we qualitatively compare its chemistry with that of MW open and GCs of similar metallicity, specifically in the range $-0.5 < \mathrm{[Fe/H]} < 0.5$, as shown in Figure \ref{Fig:comparisonOCGCs}. For this comparison, we adopt the GC catalogue of \cite{2020MNRAS.492.1641M} and the OC sample from the OCCAM project within the APOGEE survey \citep{Myers_2022}. The OCs in this sample span galactocentric distances of $6 < R_G < 12$ kpc and have ages between 1 and 7.5 Gyr, including the oldest known open cluster in the MW, NGC 6791, with an age of $8.3 \pm 0.3$ Gyr \citep{2021A&A...649A.178B}.\\
At these metallicities, the OC population is well sampled, with roughly 150 objects, while only about 9-10 GCs fall within the same [Fe/H] range. Unfortunately, the GC literature sample does not provide Ti, Cr, Mn, Co, or Ni abundances, but a meaningful comparison with the OC sample is still possible.
From this comparison (Figure \ref{Fig:comparisonOCGCs}), we see that the OCs exhibit generally flat abundance trends around $\mathrm{[X/Fe]} \approx 0$ across all considered elements, with somewhat larger dispersions for Na, Cr, and Co. In contrast, GC abundance sequences typically lie above those of the OCs. Patchick 126 often lies near the upper edge of the OC distributions and the lower edge of the GC trends. 
Also, it does not show any tight similarity with the oldest open cluster, NGC 6791, indicating that the two systems followed different chemical-evolution pathways. \\
The disc-like, orbital properties of Patchick 126, too, are typical of both classes of clusters. The number of known OCs this close to the Galactic centre is small, even though this could be due to the observational challenge of detecting low-mass objects in crowded environments, rather than being an intrinsic property of these systems. Finally, the age of Patchick~126 is very old for an OC, whereas it is consistent with in situ GCs at this metallicity.
For these reasons (and acknowledging that additional spectroscopic data are still needed), the evidence currently points toward Patchick 126 being more consistent with a GC–like nature, even though this kind of recently discovered peculiar systems points towards the fact that the traditional OC-GC distinction is more and more frail.

\end{document}